\documentclass[%
 aip,
 jmp,%
 amsmath,amssymb,
 reprint,%
]{revtex4-2}

\usepackage{graphicx}
\usepackage{subcaption} 
\usepackage{dcolumn}
\usepackage{bm}

\begin{document}

\preprint{AIP/123-QED}

\title[Study on the static detection of ICF target based on muonic X-ray sphere encoded imaging]{Study on the static detection of ICF target based on muonic X-ray sphere encoded imaging}

\author{Dikai Li}
\email{lidikai@sztu.edu.cn}
\affiliation{College of Engineering Physics, Shenzhen Technology University}

\author{Jian Yu}%
 \affiliation{College of Engineering Physics, Shenzhen Technology University}

 \author{Qian Chen}%
 \affiliation{College of Engineering Physics, Shenzhen Technology University}

\author{Chunhui Zhang}
 \affiliation{School of Nuclear Science and Technology, Lanzhou University}

\author{Xiangyu Wan}
 \affiliation{Research Center of Laser Fusion, China Academy of Engineering Physics}

\author{Leifeng Cao}
 \affiliation{College of Engineering Physics, Shenzhen Technology University}

\date{\today}

\begin{abstract}
Muon Induced X-ray Emission (MIXE) was discovered by Chinese physicist Zhang Wenyu as early as 1947, and it can conduct non-destructive elemental analysis inside samples. Research has shown that MIXE can retain the high efficiency of direct imaging while benefiting from the low noise of pinhole imaging through encoding holes. The related technology significantly improves the counting rate while maintaining imaging quality. The sphere encoding technology effectively solves the imaging blurring caused by the tilting of the encoding system, and successfully images micrometer sized X-ray sources. This paper will combine MIXE and X-ray sphere coding imaging techniques, including ball coding and zone plates, to study the method of non-destructive deep structure imaging of ICF targets and obtaining sub element distribution. This method aims to develop a new method for ICF target detection, which is particularly important for inertial confinement fusion. At the same time, this method can be used to detect and analyze materials that are difficult to penetrate or sensitive, and is expected to solve the problem of element resolution and imaging that traditional technologies cannot overcome. It will provide new methods for the future development of multiple fields such as particle physics, material science, and X-ray optics.  
\end{abstract}

\maketitle

%

\section{\label{sec:level1}Introduction}

In December 2022, the U.S. Department of Energy announced that the National Ignition Facility (NIF) at Lawrence Livermore National Laboratory (LLNL) achieved fusion ignition~\cite{Zylstra2022}. This milestone signifies that the energy produced from the fusion process exceeded the energy of the laser used to initiate it, rekindling global interest in Inertial Confinement Fusion (ICF) as a cutting-edge technology with significant implications for both fundamental physics and technological applications. 
\begin{figure}
    \centering
    \includegraphics[width=\textwidth]{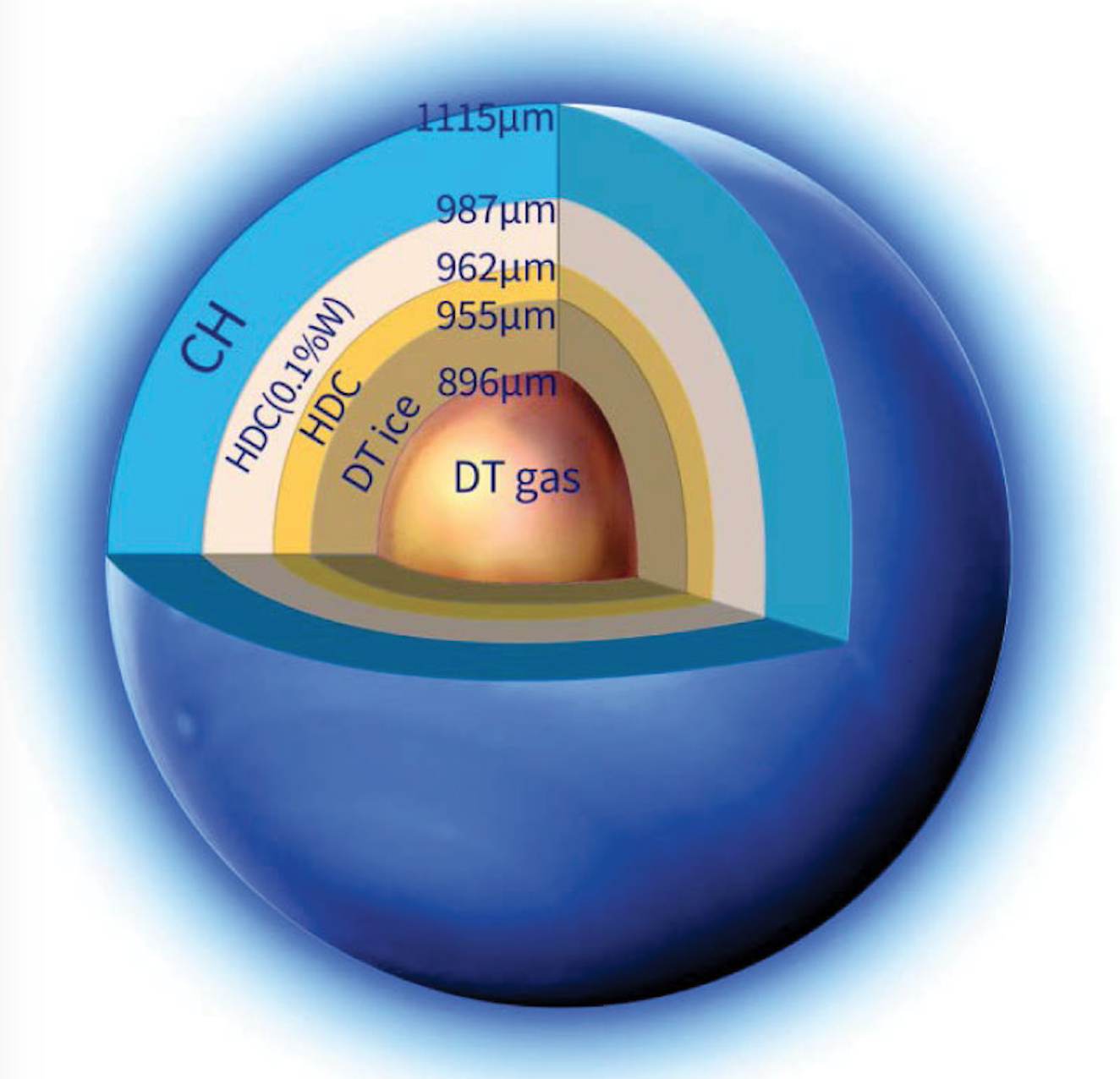}
    \caption{Caption}
    \label{fig:ICFtarget}
\end{figure}
The ICF targets are precision-engineered spherical capsules with a radius of about 1 mm, featuring concentric shell layers with thicknesses ranging from tens to hundreds of microns~\cite{Zylstra2022,PhysRevLett.126.185001}, as shown in Figure~\ref{fig:ICFtarget}. At the core of an ICF target lies a deuterium-tritium (DT) gas, surrounded by a layer of DT ice, encased further by a high-density carbon (HDC) layer, with varying designs of doped outer layers. Imperfections in the elemental distribution and cracks in the HDC and outer layers during the fabrication process can severely impact fluid dynamic stability during the fusion process. Therefore, pre-fusion static diagnostics, including precise imaging of the target capsule shape and assessing the uniformity of elemental distribution, are crucial. Current non-destructive, static internal structure examination techniques primarily involve X-ray imaging and laser differential confocal measurement, which provide imaging of the internal structure along with density distribution and layer thickness information~\cite{2012-6-068703,10.1063/1.5085863,10.11884/HPLPB202032.200136}. However, these techniques currently struggle to precisely determine the elemental distribution inside ICF targets. This paper proposes to explore Muon-Induced X-ray Emission (MIXE) as a technique for high-resolution, non-destructive, deep-layer elemental imaging of ICF targets.

Muons were initially discovered in 1936 by Carl D. Anderson and Seth Neddermeyer during cosmic ray experiments, are leptons carrying a unit electric charge. The muon ($\mu^{\pm}$) has a mass approximately 207 times that of an electron and an average lifespan of 2.2 microseconds. Its longevity is only surpassed by that of neutrons among unstable particles, which constitutes a significant advantage in its applications. Electrons and muons are the only final-state leptons that can be directly detected in experiments. Due to these distinctive properties, muons play a crucial role in both fundamental physics and applied technologies~\cite{2020-49-10-001,2021-50-4-006,2021-50-4-007,2021-50-4-008}.

As early as 1947, the Chinese physicist Zhang Wenyu discovered that negative muons ($\mu^{-}$) can be captured by the Coulomb field of atomic nuclei to form muonic atoms, subsequently releasing X-rays during cascade de-excitation~\cite{RevModPhys.21.166}. Muon Induced X-ray Emission (MIXE) produces energy related to the captured atom, similar to characteristic X-rays of electrons~\cite{2021-50-4-007}. Due to the larger rest mass of the negative muon compared to that of the electron, the emission energy of muon X-rays (0.2-6 MeV) is also higher than that of electron X-rays~\cite{HILLIER2016203}. Consequently, MIXE can easily exit large samples without significant photon self-absorption. This is particularly advantageous for samples containing atoms of low atomic numbers, where high-energy resolution semiconductor detectors can be employed to determine elemental compositions~\cite{Ninomiya_2010}. The momentum of negative muon beams provided by accelerator-based muon sources is adjustable, allowing for depth-dependent elemental analysis within materials. MIXE has been successfully applied in studies of archaeological objects~\cite{Clemenza2019}, biological sciences~\cite{Daniel1987}, extraterrestrial material detection~\cite{Terada2017}, lithium batteries~\cite{doi:10.1021/acs.analchem.0c00370}, among other applications~\cite{HILLIER2016203,nano10071260,Shimada-Takaura2021}.

There have been numerous studies on Muon Induced X-ray Emission (MIXE). The SIN facility in Switzerland (now PSI), Canada's TRIUMF, and the LANMF muon source in the USA have conducted MIXE elemental analyses on biomaterials~\cite{doi:10.1021/ac50023a015,10.1259/0007-1285-68-816-1325}. The RIKEN-RAL muon source at the UK's ISIS, the MUSE muon source at Japan's J-PARC, and the RCNP/MuSIC muon source at Osaka University in Japan are all actively engaged in MIXE elemental analysis experiments~\cite{Terada2017,heritage2010028,doi:10.1021/acs.analchem.5b01169}.

The spatial distribution of elements obtained via Muon Induced X-ray Emission (MIXE) can be imaged using pixel detectors. For instance, Hillier placed a CdTe pixel detector 4 mm behind the sample at ISIS using the High Energy X-ray Imaging Technology (HEXITEC) for direct imaging~\cite{doi:10.7566/JPSCP.21.011042}. Katsurakawa developed an imaging system based on a CdTe double-sided strip detector (CdTe-DSD) and conducted pinhole imaging~\cite{KATSURAGAWA2018140}. Chiu utilized a pinhole at J-PARC, rotated the sample, and employed the Maximum Likelihood Expectation Maximization algorithm to reconstruct the sample's three-dimensional distribution~\cite{Chiu2022}.

Previously, China did not have a muon source. The second phase of the China Spallation Neutron Source (CSNS) is currently constructing a muon source for science, technology, and industry (Muon station for sciEnce, technoLOgy and inDustrY, MELODY)~\cite{Bao_2023}. A high-brightness muon source will also be built at the China Initiative Accelerator Driven System (CiADS)~\cite{PhysRevAccelBeams.27.023403}. Both MELODY II and the muon source upgrade plans at CiADS will provide sufficient brightness and beam time for MIXE applications. At MELODY II, the negative muon beam for MIXE will have a maximum momentum of 30 MeV/c with a repetition rate of 5 Hz. Compared to ISIS (40 Hz)~\cite{Eaton1992} and J-PARC (25 Hz)~\cite{MIYAKE200922}, MELODY's repetition rate is still lower, which will substantially limit the counting rate for future MIXE applications.

To address this issue, Pan employed an improved uniformly redundant array (MURA)~\cite{Gottesman:89} coded aperture imaging technique in MIXE applications~\cite{LIN2022166783}. They conducted a feasibility study on coded imaging technology for elemental analysis using MIXE, based on Geant4 simulations~\cite{LIN2022166783}, this approach not only retains the high efficiency of direct imaging~\cite{doi:10.7566/JPSCP.21.011042} but also benefits from the low noise of pinhole imaging~\cite{KATSURAGAWA2018140}. Furthermore, preliminary experiments using gamma rays emitted from a $^{22}$Na source for coded aperture imaging were reported at MELODY2023 held in November 2023 at CSNS; the results demonstrated a significant increase in counting rate while maintaining image quality.

Clearly, Muon Induced X-ray Emission (MIXE) provides characteristic X-rays for elemental analysis, offering detailed information about the distribution of different elements within a sample. The larger mass of muons and the adjustability of muon beam momentum produced by accelerators facilitate deep penetration into target layers of the sample. Given that different shell layers of ICF targets are composed of various materials, employing MIXE combined with coded imaging technology for the inspection of ICF targets could offer higher interface sensitivity than traditional X-ray techniques due to MIXE's capability for elemental resolution.

When combined with coded imaging technology, Muon Induced X-ray Emission (MIXE) achieves better imaging quality and higher counting rates~\cite{doi:10.7566/JPSCP.21.011042,KATSURAGAWA2018140,Chiu2022,LIN2022166783}, which is particularly relevant given the low repetition rate of future muon sources in China. However, the spatial resolution of current MIXE combined with MURA coding is typically on the millimeter scale~\cite{LIN2022166783}, which does not meet the precise inspection requirements for ICF target shells that are tens of micrometers thick. A published work by our team utilized a spherical coding imaging technique to accurately characterize X-ray source sizes in the tens of micrometers, effectively mitigating the blurring caused by tilting common in general coded imaging systems. This was further enhanced by using the Richardson-Lucy algorithm to iteratively deconvolve and reconstruct the spatial information of the X-ray source~\cite{10.1063/5.0180056}. According to references~\cite{10.1063/5.0130689,10.1063/1.4959161}, with the appropriate choice of detector, object distance, and magnification, the spatial resolution of the spherical coded imaging system can reach the micrometer level.

This paper combines Muon Induced X-ray Emission (MIXE) with spherical coded imaging technology to explore a method for non-destructive, high-resolution, deep-layer elemental imaging of ICF targets. Focusing on the concentric shell structure and elemental distribution of the target capsules, muon beam irradiation of ICF targets is simulated using Geant4. By adjusting the muon beam momentum, muons penetrate to the target layers requiring examination, such as the high-density carbon (HDC) layers, generating muon-induced X-rays. The characteristic X-rays of the target elements, containing spatial distribution information, pass through the spherical coding system to form an encoded MIXE image of the target capsule. The image, containing deep-layer elemental distribution information of the target, is then reconstructed using decoding algorithms. Following the methods introduced in this paper, future guidance can be provided for setting up X-ray spherical coded imaging experiments using radiation sources to validate the X-ray spherical coding system and reconstruction algorithms discussed herein, thereby analogously studying their performance in ICF target MIXE. This can also guide the implementation of high spatial resolution elemental analysis experiments on suitable muon sources using MIXE combined with spherical coding technology.

With the successful ignition at NIF, the urgency of ICF research in China is undeniable, and the preparation and inspection of ICF target capsules are of critical importance. Against the backdrop of the construction of China's first muon source, muon-related sciences are poised for vigorous development. By integrating MIXE with spherical coding, this paper investigates elemental analysis techniques for ICF targets, aiming to enhance the spatial resolution of elemental imaging of ICF target capsules to the micrometer level, thereby making significant contributions to the preparation and experimental quality control of ICF targets.

\section{Methods}

\subsection{Geant4 simulation}

\subsubsection{The projection of a sphere with uniform volumetric density onto a two-dimensional plane}
When muons uniformly irradiate the shell or sphere of an ICF target ball, they transform the sample into a uniform glowing shell or spherical X-ray source through the MIXE process. In the experiments, a coded sphere is placed behind the ICF target, followed by a detector to capture the emitted X-rays. Therefore, before simulation, it is necessary to understand the surface density distribution obtained by projecting a shell or sphere with uniform volumetric density onto a two-dimensional plane, in order to preliminarily verify whether the distribution of X-rays at the generation site meets the expected criteria.

\begin{figure}
    \centering
    \begin{subfigure}[b]{0.45\textwidth} 
        \centering
        \includegraphics[width=\textwidth]{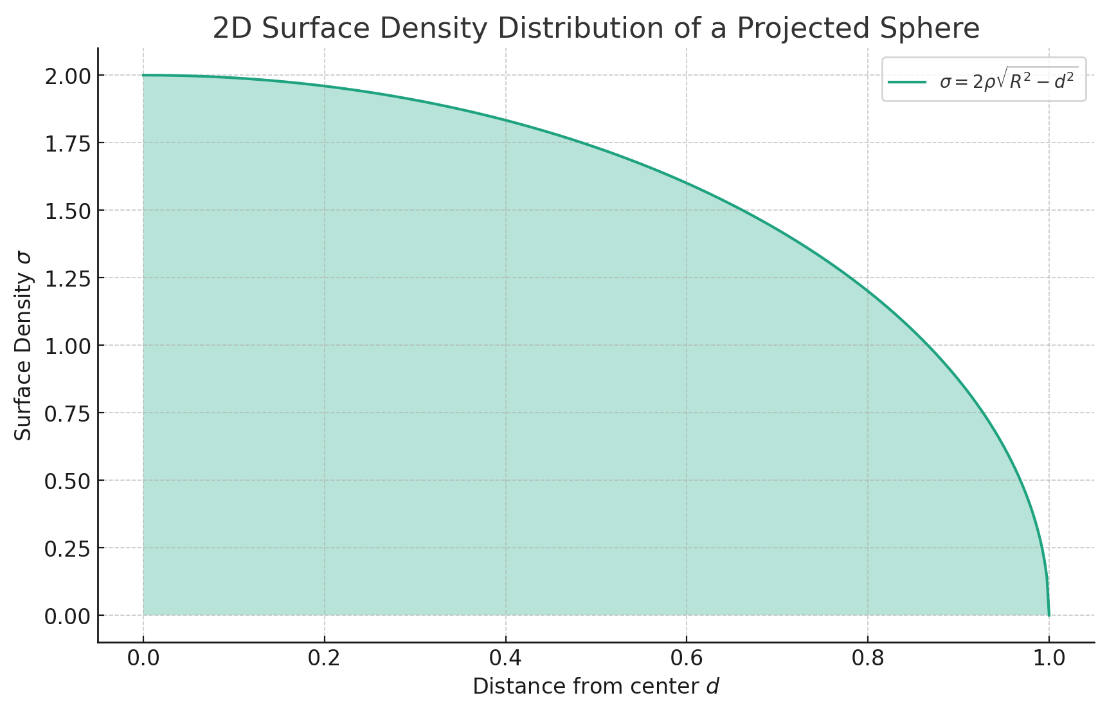} 
        \caption{}
        \label{fig:1a}
    \end{subfigure}
    \hfill 
    \begin{subfigure}[b]{0.45\textwidth} 
        \centering
        \includegraphics[width=\textwidth]{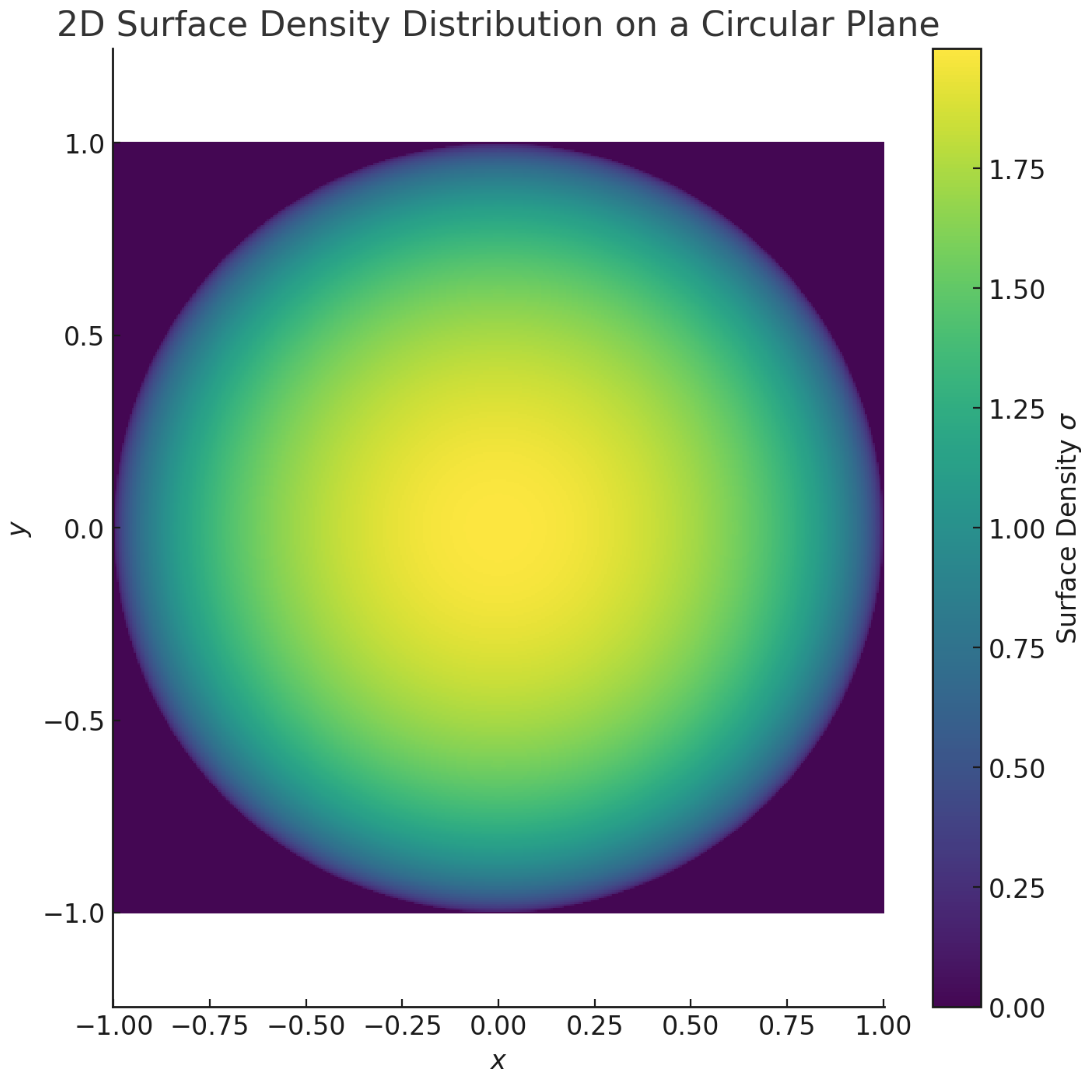} 
        \caption{}
        \label{fig:1b}
    \end{subfigure}
    \caption{2D Projection of a Spherical Volumetric Density onto a One-Dimensional Plane.}
    \label{fig:1}
\end{figure}

The formula for the surface density distribution when a uniformly distributed sphere is projected onto a two-dimensional plane is as follows,
\begin{equation}
    \sigma = 2\rho \sqrt{R^2-r^2}
\end{equation}
Here, $\sigma$ represents the surface density of the projection on the two-dimensional plane,$\rho$ represents the volumetric density of the sphere, R is the radius of the sphere, and r is the distance from a point on the two-dimensional plane to the projection of the sphere's center on that plane. Based on this formula, a graph of the two-dimensional surface density distribution can be drawn as Figure~\ref{fig:1}.
The left image illustrates how the surface density gradually decreases from its maximum at the center towards the edge. The right image uses color to display the changes in two-dimensional surface density.

To obtain the projection of a spherical shell with thickness on a two-dimensional plane, consider removing a smaller-radius sphere from within a larger-radius sphere. This provides the formula for the surface density distribution of the spherical shell projection,
\begin{equation}
    \sigma = 2\rho(\sqrt{R_2^2-r^2}-\sqrt{R_1^2-r^2})
\end{equation}
Here, $R_1$ is the radius of the smaller sphere, and $R_2$ is the radius of the larger sphere. 
\begin{figure}
    \centering
    \begin{subfigure}[b]{0.45\textwidth} 
        \centering
        \includegraphics[width=\textwidth]{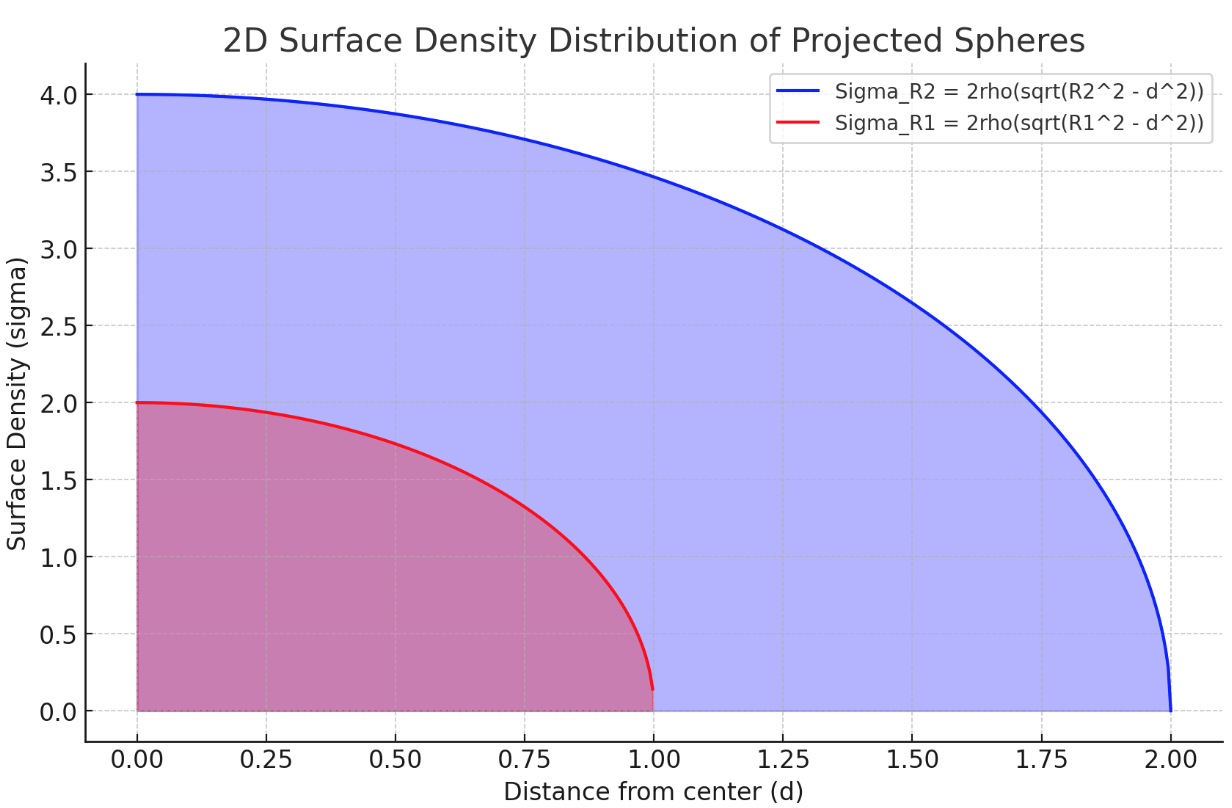} 
        \caption{}
        \label{fig:2a}
    \end{subfigure}
    \hfill 
    \begin{subfigure}[b]{0.45\textwidth} 
        \centering
        \includegraphics[width=\textwidth]{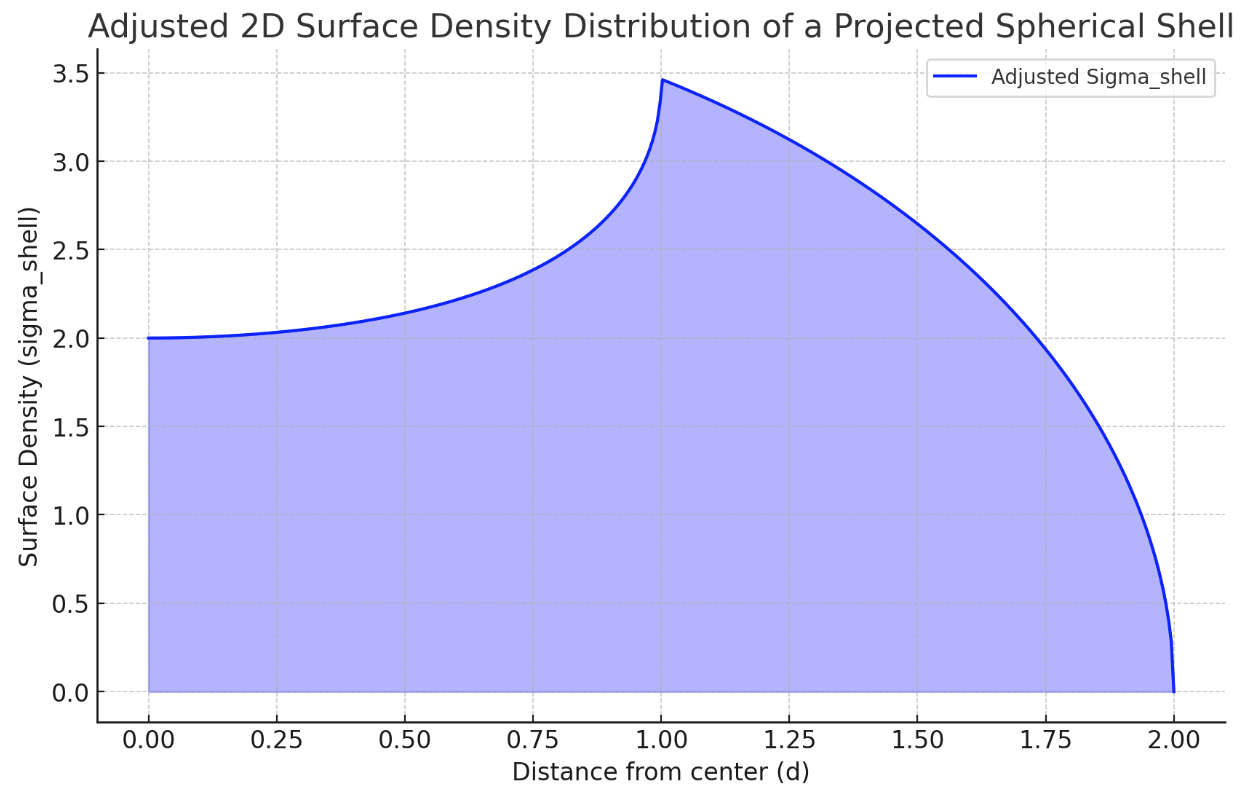} 
        \caption{}
        \label{fig:2b}
    \end{subfigure}
    \hfill 
    \begin{subfigure}[b]{0.45\textwidth} 
        \centering
        \includegraphics[width=\textwidth]{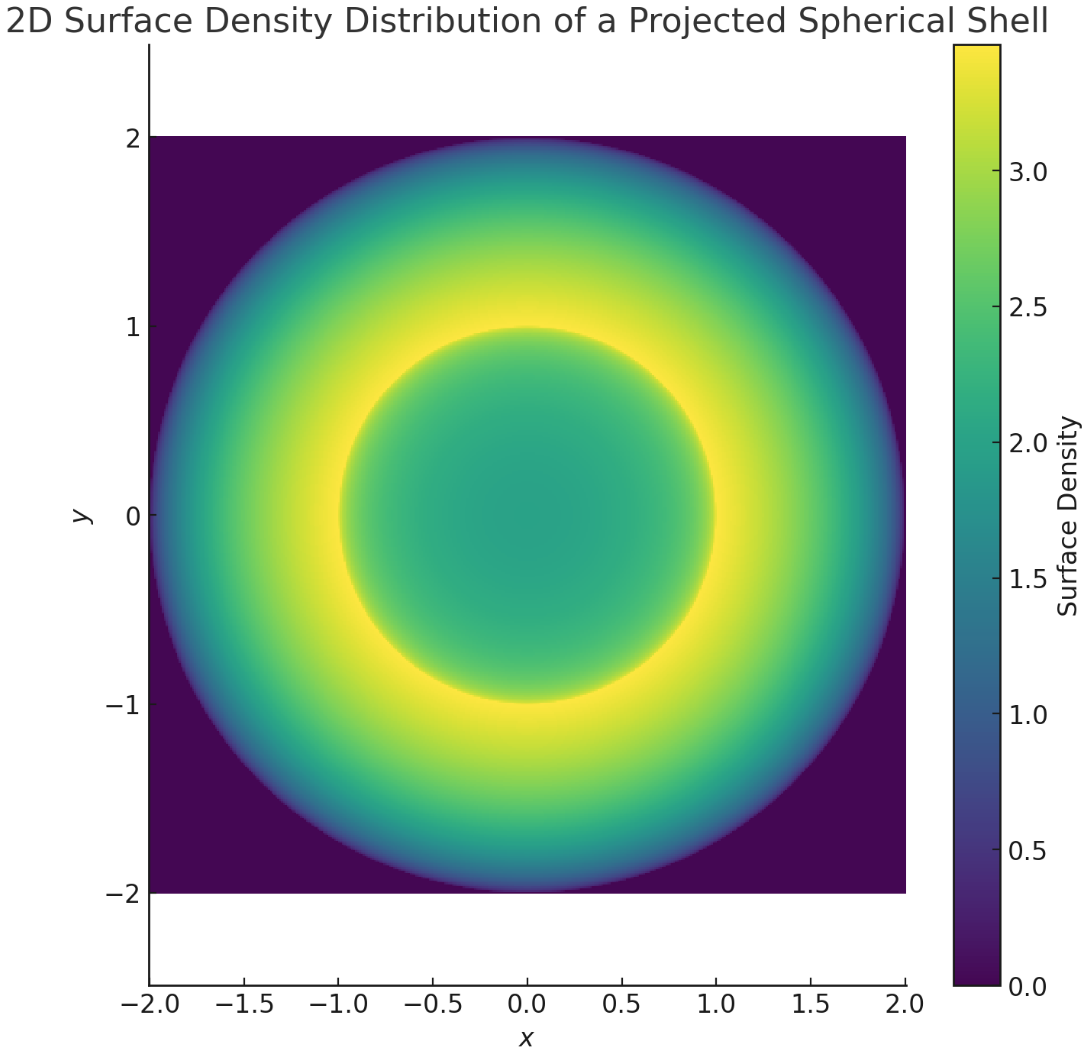} 
        \caption{}
        \label{fig:2c}
    \end{subfigure}
    \caption{a.One-Dimensional Projection of Two Spherical Densities onto a Two-Dimensional Plane,b.Projection of a Thick Spherical Shell's Density onto a One-Dimensional Two-Dimensional Plane,c.Two-Dimensional Projection of a Thick Spherical Shell's Density onto a Two-Dimensional Plane.}
    \label{fig:2}
\end{figure}
Figure~\ref{fig:2a} displays the surface density distributions of the two spherical projections separately, Figure~\ref{fig:2b} shows the final surface density distribution obtained by subtracting the effect of the smaller sphere from the larger sphere, and Figure~\ref{fig:2c} uses color to indicate the magnitude of the surface density.

Now, incorporating the data from an ICF target pellet, such as for a high-density carbon (HDC) spherical shell with an inner radius $R_1=955\mu m$ and an outer radius $R_2=962\mu m$, the formula can be used to plot the surface density distribution of a uniformly distributed HDC spherical shell projected onto a two-dimensional plane as shown in the figures. 
\begin{figure}
    \centering
    \begin{subfigure}[b]{0.45\textwidth} 
        \centering
        \includegraphics[width=\textwidth]{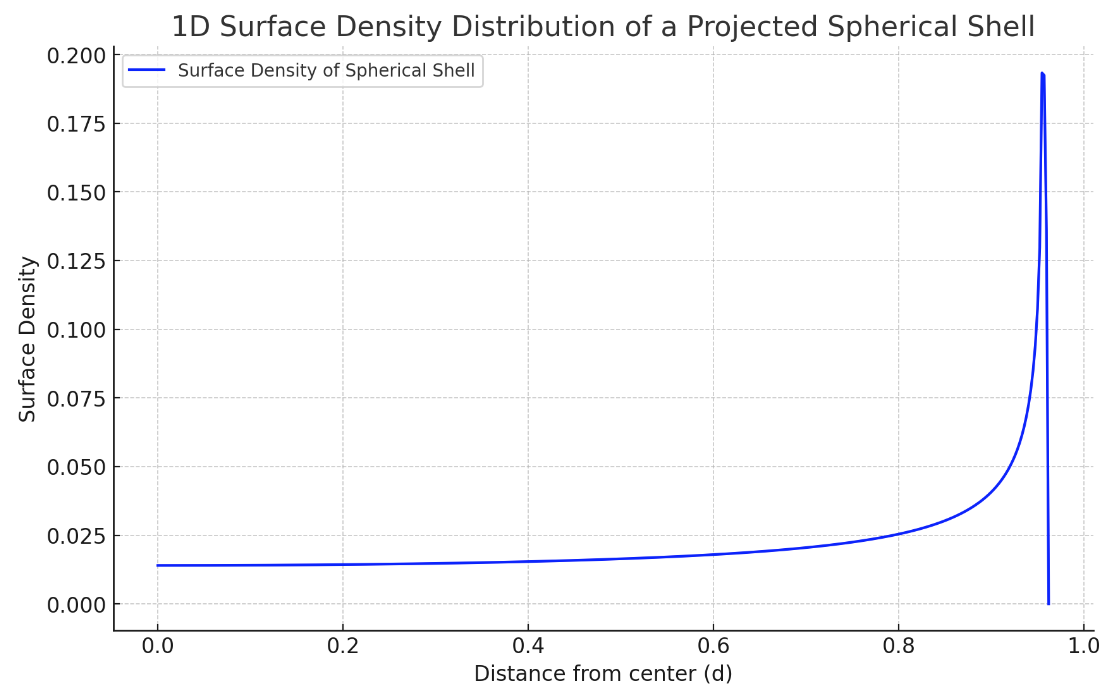} 
        \caption{}
        \label{fig:3a}
    \end{subfigure}
    \hfill 
    \begin{subfigure}[b]{0.45\textwidth} 
        \centering
        \includegraphics[width=\textwidth]{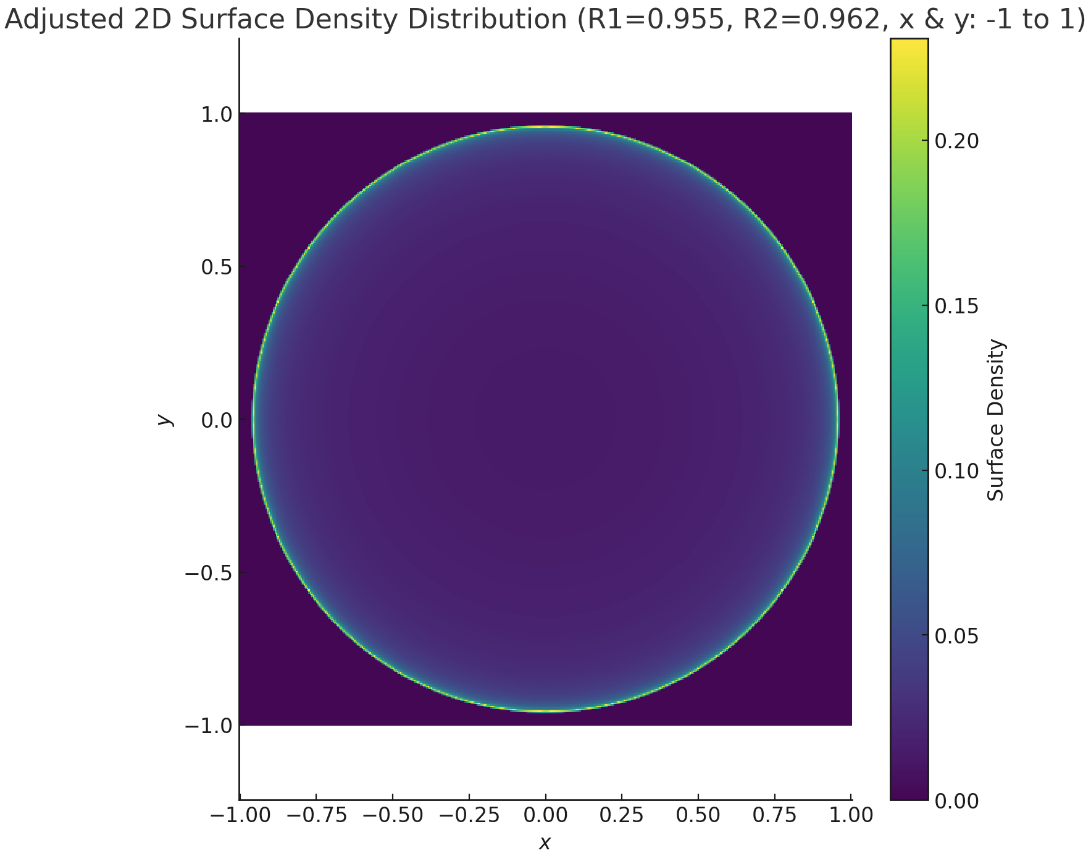} 
        \caption{}
        \label{fig:3b}
    \end{subfigure}
    \caption{a.One-Dimensional Projection of a Thick ICF Spherical Shell's Density onto a Two-Dimensional Plane,b.Two-Dimensional Projection of a Thick ICF Spherical Shell's Density onto a Two-Dimensional Plane}
    \label{fig:3}
\end{figure}
From Figure~\ref{fig:3a}, it can be observed that the surface density is very low in most regions extending outward from the center of the sphere, until it sharply increases near the inner radius of the shell, forming a steep peak. Such characteristics can be used in simulations to determine if the process of muon-induced X-ray production in the spherical shell has been correctly generated, and in decoding to ascertain if the reconstructed image accurately reflects the shell's features. In experiments, the obtained images can be compared with the features in Figure 3a; if significant discrepancies are observed, it suggests that the sampled shell is not uniformly distributed, indicating potential flaws or defects. This provides critical evidence for the inspection of ICF target pellets. It is important to note that this spherical shell model is only applicable in cases where the space outside the shell is a vacuum. When materials fill the area inside the inner radius or beyond the outer radius of the shell, additional considerations must be taken into account.

\subsubsection{Geant4 simulation on MIXE process on ICF target}
The study will focus on the Muon Induced X-ray Emission (MIXE) process involving the interaction between muons and ICF target capsules. The core of an ICF target capsule consists of deuterium-tritium (DT) gas, surrounded by a DT ice layer, further encased by high-density carbon (HDC), hydrocarbon polymers (CH), and various other designed layers. This paper will establish the geometry of the ICF target capsules in Geant4, set up the elemental distributions of each layer, and preset hypothetical irregularities such as cracks, followed by directing a muon beam to irradiate the target. The study will examine the momentum dependency of the muon beam as it penetrates into the target capsule and further investigate the X-rays emitted under different positions and elemental distributions, particularly within the inner walls of the HDC layer, which directly affects the uniformity of the DT fuel.

Utilizing Geant4 version 11.2.1, this paper initially established a structural model of the ICF target capsule shell layers based on Figure 1 (left) of literature~\cite{PhysRevLett.126.185001}. The model consists of a DT gas layer with a density of 0.44 mg/cm³ filled with a deuterium-tritium (1:1) mixture, a DT ice layer with a density of 0.255 g/cm³ also with a deuterium-tritium (1:1) mixture, an HDC layer with a density of 3.48 g/cm³ composed of carbon, an HDC (0.1\%W) layer with a density of 3.52 g/cm³ carbon doped with 0.1\% tungsten, and a CH layer filled with a carbon-hydrogen (1:1) polymer at a density of 1.05 g/cm³. The thicknesses of these layers are set according to the radius shown in Figure 1 (right). Based on the negative muon beam parameters described at the MELODY2023 conference and in literature~\cite{Bao_2023}, the simulation involves irradiating the ICF target capsule with muons at a maximum momentum of 30 MeV/c. The QBBC physics list was chosen to simulate processes such as nuclear capture of negative muons, de-excitation of muonic atoms emitting X-rays, muon decay, and various electromagnetic interactions within the sample. The simulations employ CdTe and CdZnTe pixel detectors to record the energy, location, and direction of X-ray hits. This data will later be used to reconstruct the elemental distribution within the target capsule.

\begin{figure}
    \centering
    \begin{subfigure}[b]{0.45\textwidth} 
        \centering
        \includegraphics[width=\textwidth]{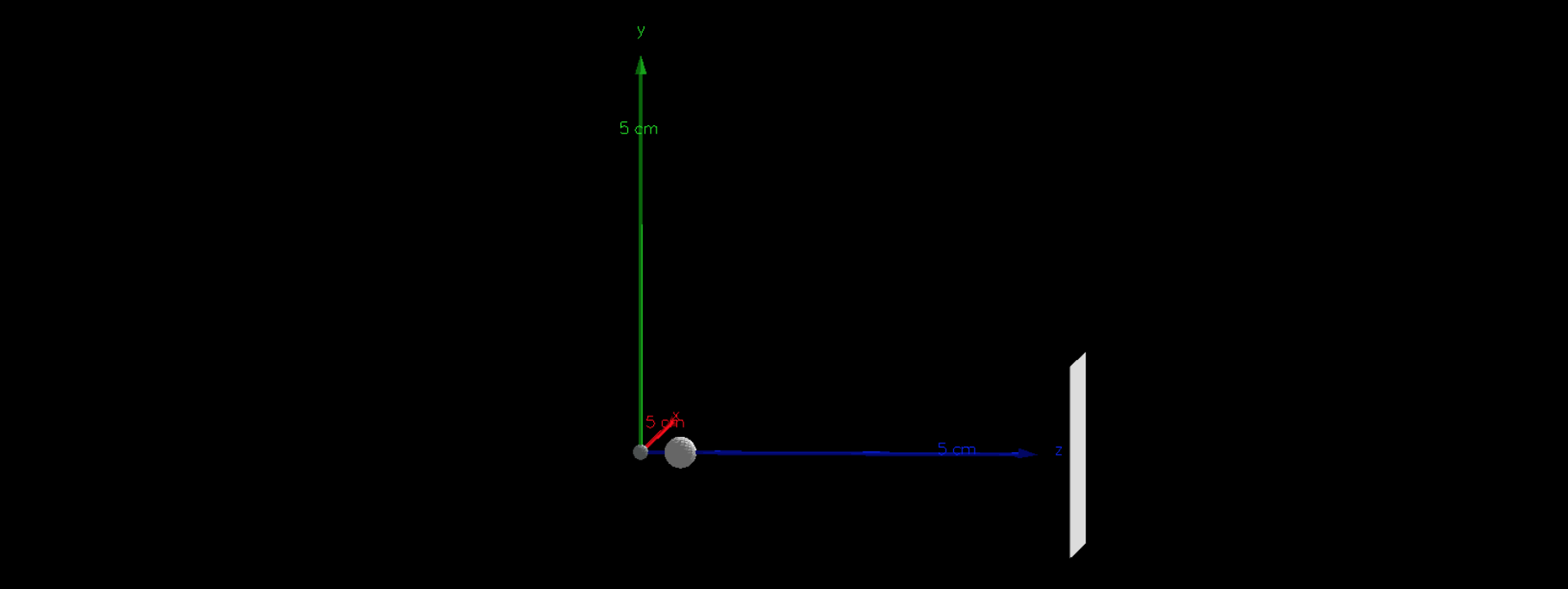} 
        \caption{}
        \label{fig:4a}
    \end{subfigure}
    \hfill 
    \begin{subfigure}[b]{0.45\textwidth} 
        \centering
        \includegraphics[width=\textwidth]{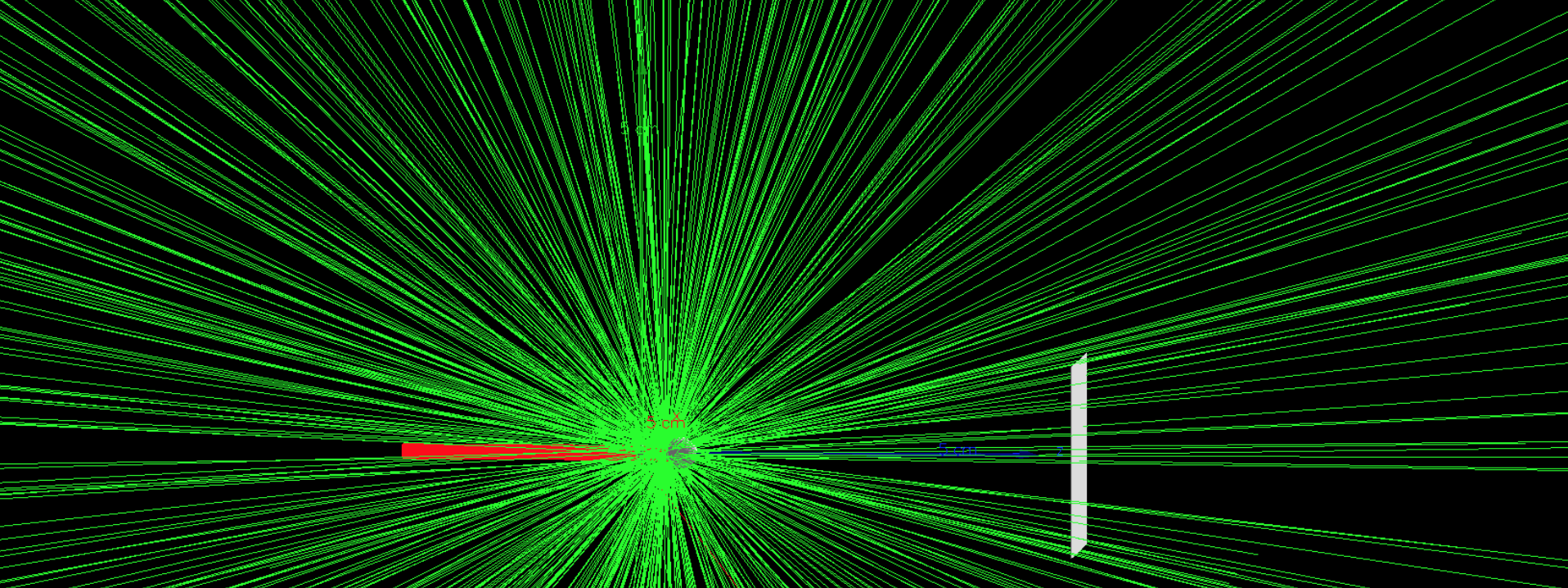} 
        \caption{}
        \label{fig:4b}
    \end{subfigure}
    \caption{Geometry in Geant4}
    \label{fig:3}
\end{figure}
Figure~\ref{fig:4a}: Geant4 simulation geometry showing the smallest sphere on the left as the HDC shell of the ICF target, with a lead sphere placed next to it and a flat panel detector on the far right. Figure~\ref{fig:4b}: The red beam lines on the left represent the incident muon beam, generating isotropically emitted X-rays (green lines) at the ICF target.

\begin{figure}
    \centering
    \includegraphics[width=\textwidth]{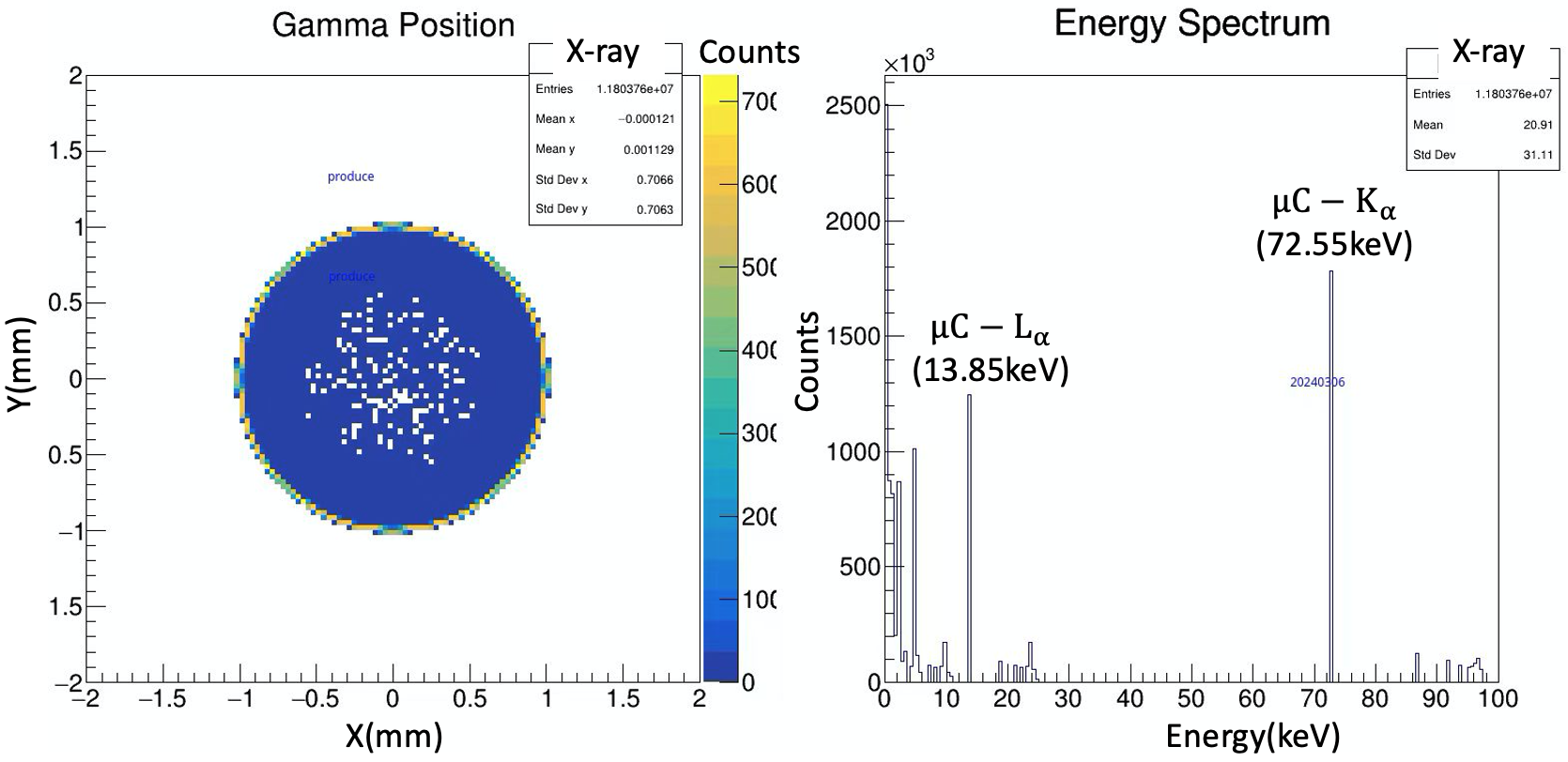}
    \caption{X-ray produced on ICF target, induced by muon beam}
    \label{fig:5}
\end{figure}
Figure 5~\ref{fig:5} left: Spatial distribution of the generated X-rays, consistent with the spherical shell projection shown in Figure~\ref{fig:3}. Figure 5~\ref{fig:5} right: Energy spectrum of the generated X-rays, displaying a distinct peak corresponding to carbon.

\subsection{Sphere encoding process}
\begin{figure}
    \centering
    \includegraphics[width=\textwidth]{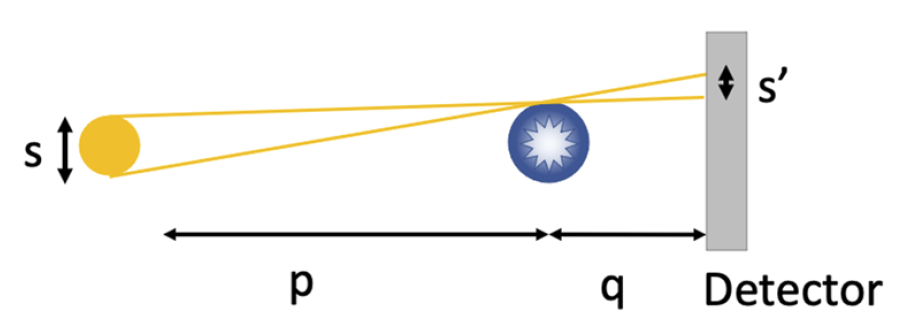}
    \caption{Sphere coding system}
    \label{fig:6}
\end{figure}

According to reference~\cite{10.1063/5.0130689,10.1063/1.4959161}, a sphere coding system spatial resolution can be estimated as
\begin{equation}
    h=\sqrt{(\frac{e}{M})^2+p\lambda+L^2+(\Delta d)^2}
\end{equation}
where e is the detector resolution, M is the magnification, $\lambda$ is the x-ray wavelength, L is a term related to the penetration length through the edge of the aperture,
and $\Delta d$ is a measure of the circularity of the aperture. The first term takes into account the detector resolution, which can, in principle,
be made negligible with a large enough magnification. The L and $\Delta d$ terms can be kept small compared to the diffraction term. For the 77.25 keV $\mu$C-$K_\alpha$ X-ray, $\lambda=1.709\times 10^{-11}$m.

\begin{figure}
    \centering
    \includegraphics[width=\textwidth]{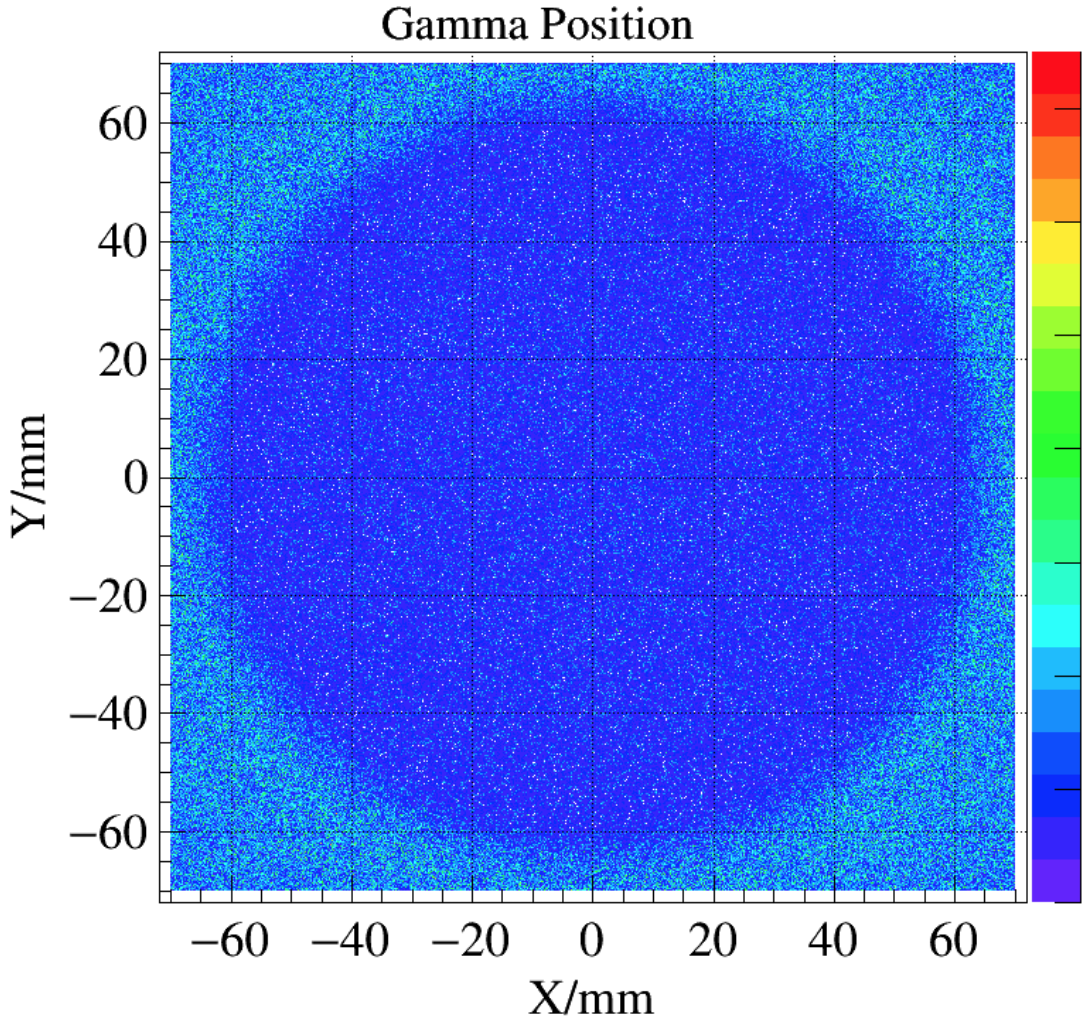}
    \caption{Coded image}
    \label{fig:7}
\end{figure}
Figure~\ref{fig:7} is coded image.

\subsection{Reconstruction}
\begin{figure}
    \centering
    \includegraphics[width=\textwidth]{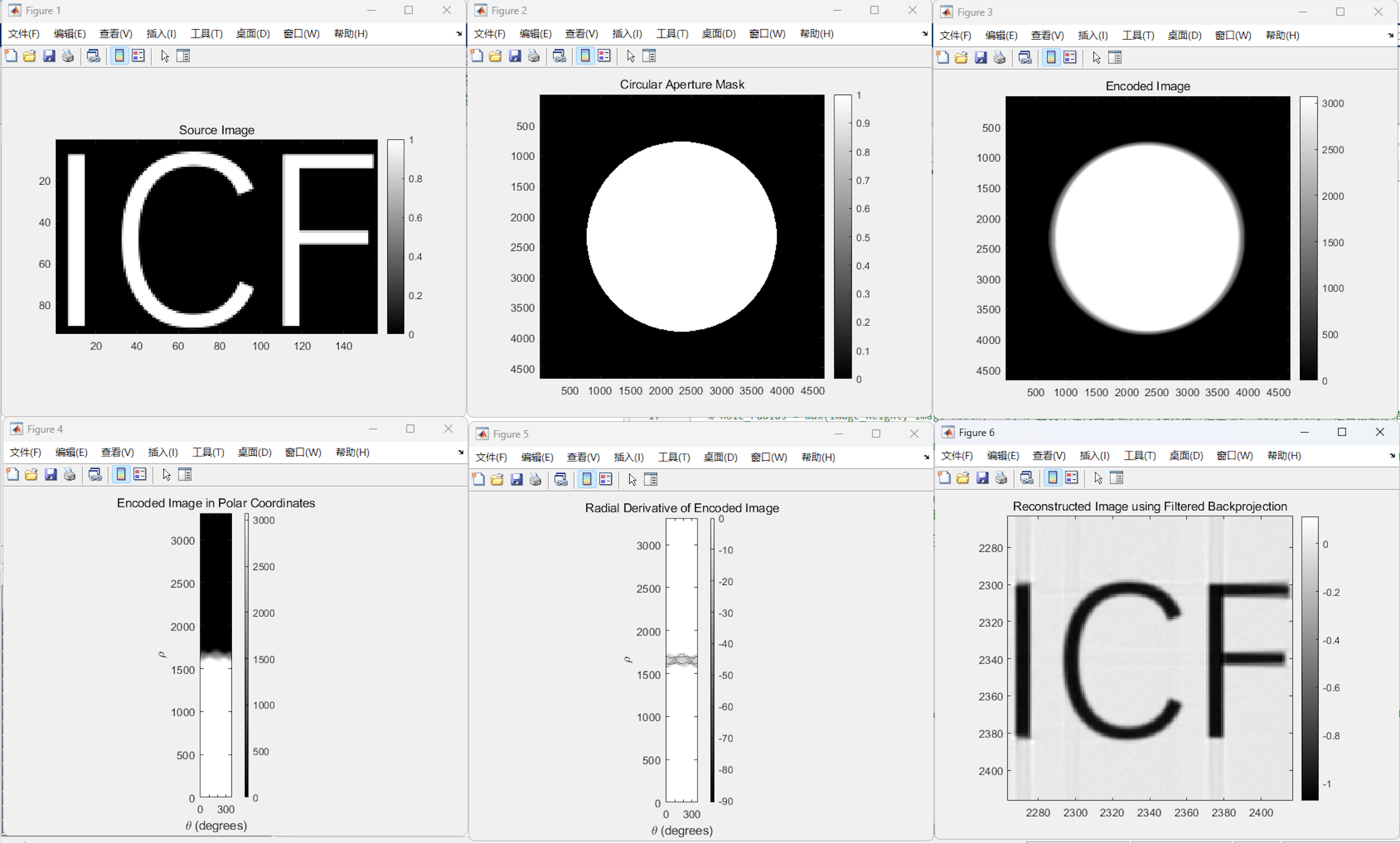}
    \caption{Caption}
    \label{fig:sphere_hole_decdoded}
\end{figure}

We used the inverse Radon method in CT algorithm for reconstruction. Figure \ref{fig:sphere_hole_decdoded} shows the object encoded through a larger transparent circular hole, which forms an encoded image on the detector. Then, through the inverse Radon decoding process, the original image of the object is reconstructed.

\begin{figure}
    \centering
    \includegraphics[width=\textwidth]{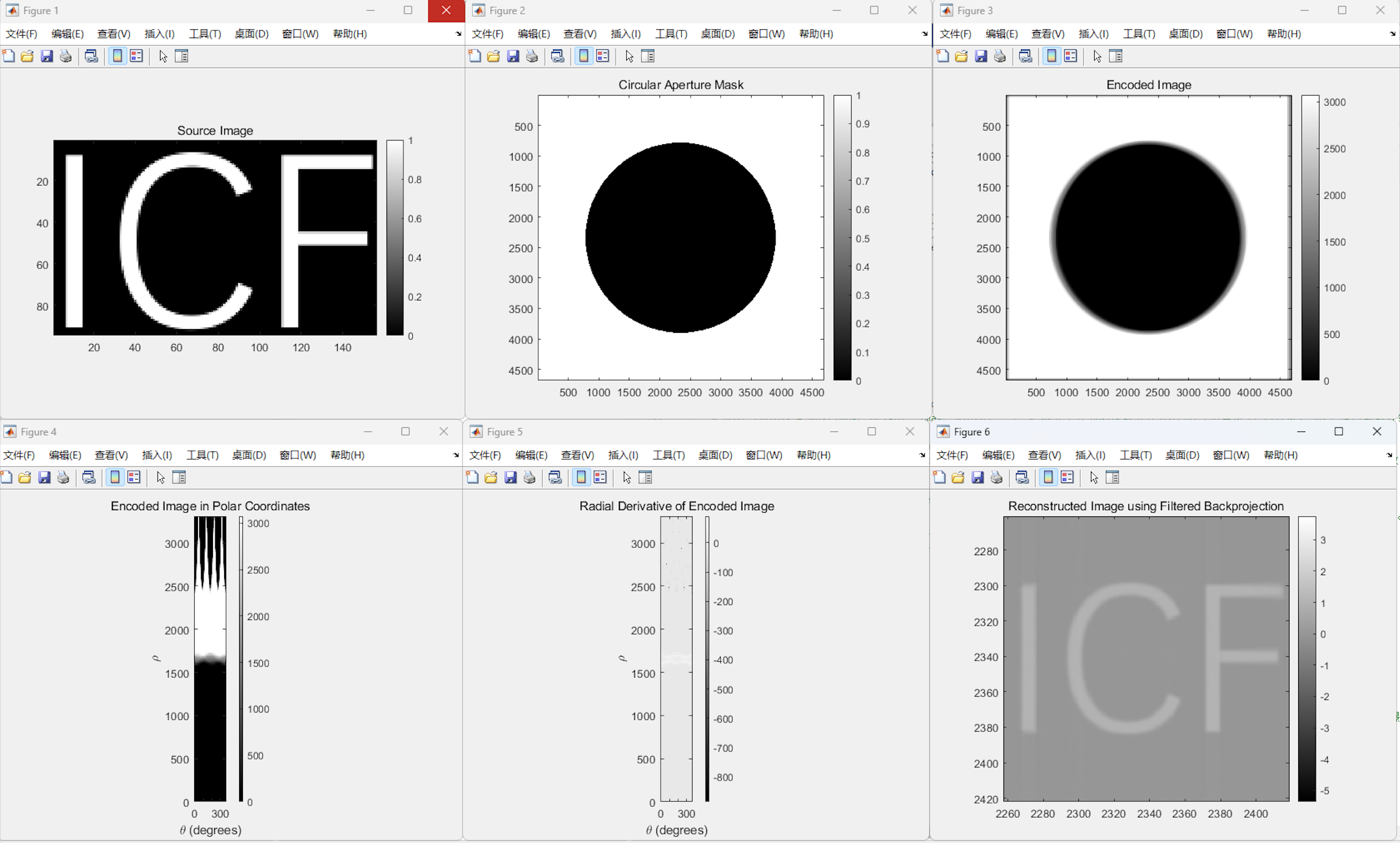}
    \caption{Caption}
    \label{fig:sphere_plate_decdoded}
\end{figure}

Figure \ref{fig:sphere_plate_decdoded} shows the object encoded by a larger opaque disk, which reaches the detector to form an encoded image. Then, through the inverse Radon decoding process, the original image of the object is reconstructed.

In the experiment, an opaque sphere will replace the opaque disk, with the same optical properties. However, the installation and adjustment of the sphere will be more convenient, and it can easily overcome the problems of circular holes and non-circular disks caused by tilting.

The encoded image is then subjected to a decoding reconstruction process. Initially, the encoded image data is imported into MATLAB and stored as a grayscale matrix. 

Subsequent deconvolution is performed using Wiener filtering and the Richardson-Lucy (R-L) algorithm. This step requires the encoded sphere matrix, which is generated using the formula for the two-dimensional projection of the surface density distribution of a spherical body. The final output is the reconstructed image as shown in Figure~\ref{}.

\section{Near future work}
\begin{itemize}
\item Adjusting the muon beam's kinetic energy settings to determine the optimal muon incident energy that maximizes the induction of characteristic X-rays from target atoms in ICF target pellets.
\item Conducting experiments where muon irradiation and subsequent coding processes are applied to ICF target pellets with overlapping shell layers, and decoding images based on energy-selected X-rays characteristic of specific elements via detectors.
\item Implementing different decoding algorithms and further optimizing parameters such as the noise-to-signal ratio in iterative Wiener filtering and the number of iterations in Richardson-Lucy (R-L) algorithms to achieve improved reconstruction images.
\item Further analysis of the decoded reconstruction images to establish metrics for imaging quality, resolution, and uniformity. 
\end{itemize}
These studies will extend the current findings and enhance our understanding of the imaging and analytical capabilities of the muon-induced X-ray emission technique in complex layered targets.

\section{Summary and discussion}
This study demonstrates a combined application of Muon Induced X-ray Emission (MIXE) and spherical coded imaging technology for advanced, non-destructive imaging of ICF targets, emphasizing high-resolution and deep-layer elemental analysis. Utilizing Geant4 simulations, the process involves adjusting muon beam momentum to achieve targeted penetration into specific layers such as HDC, thereby inducing characteristic X-rays. These X-rays are encoded by a spherical system to produce an image that encapsulates the elemental distribution within the target’s deeper layers. Decoding algorithms are then employed to reconstruct these images, providing a comprehensive view of the internal elemental structure. The methodologies outlined offer a foundation for future experimental setups to validate the spherical coding system and enhance the understanding of MIXE applications in elemental analysis at high spatial resolutions.




\nocite{*}
\bibliography{aipsamp}

\providecommand{\noopsort}[1]{}\providecommand{\singleletter}[1]{#1}%
\begin{thebibliography}{34}%
\makeatletter
\providecommand \@ifxundefined [1]{%
 \@ifx{#1\undefined}
}%
\providecommand \@ifnum [1]{%
 \ifnum #1\expandafter \@firstoftwo
 \else \expandafter \@secondoftwo
 \fi
}%
\providecommand \@ifx [1]{%
 \ifx #1\expandafter \@firstoftwo
 \else \expandafter \@secondoftwo
 \fi
}%
\providecommand \natexlab [1]{#1}%
\providecommand \enquote  [1]{``#1''}%
\providecommand \bibnamefont  [1]{#1}%
\providecommand \bibfnamefont [1]{#1}%
\providecommand \citenamefont [1]{#1}%
\providecommand \href@noop [0]{\@secondoftwo}%
\providecommand \href [0]{\begingroup \@sanitize@url \@href}%
\providecommand \@href[1]{\@@startlink{#1}\@@href}%
\providecommand \@@href[1]{\endgroup#1\@@endlink}%
\providecommand \@sanitize@url [0]{\catcode `\\12\catcode `\$12\catcode `\&12\catcode `\#12\catcode `\^12\catcode `\_12\catcode `\%12\relax}%
\providecommand \@@startlink[1]{}%
\providecommand \@@endlink[0]{}%
\providecommand \url  [0]{\begingroup\@sanitize@url \@url }%
\providecommand \@url [1]{\endgroup\@href {#1}{\urlprefix }}%
\providecommand \urlprefix  [0]{URL }%
\providecommand \Eprint [0]{\href }%
\providecommand \doibase [0]{https://doi.org/}%
\providecommand \selectlanguage [0]{\@gobble}%
\providecommand \bibinfo  [0]{\@secondoftwo}%
\providecommand \bibfield  [0]{\@secondoftwo}%
\providecommand \translation [1]{[#1]}%
\providecommand \BibitemOpen [0]{}%
\providecommand \bibitemStop [0]{}%
\providecommand \bibitemNoStop [0]{.\EOS\space}%
\providecommand \EOS [0]{\spacefactor3000\relax}%
\providecommand \BibitemShut  [1]{\csname bibitem#1\endcsname}%
\let\auto@bib@innerbib\@empty
\bibitem [{\citenamefont {Zylstra}\ \emph {et~al.}(2022)\citenamefont {Zylstra}, \citenamefont {Hurricane}, \citenamefont {Callahan}, \citenamefont {Kritcher}, \citenamefont {Ralph}, \citenamefont {Robey}, \citenamefont {Ross}, \citenamefont {Young}, \citenamefont {Baker}, \citenamefont {Casey}, \citenamefont {D{\"o}ppner}, \citenamefont {Divol}, \citenamefont {Hohenberger}, \citenamefont {Le~Pape}, \citenamefont {Pak}, \citenamefont {Patel}, \citenamefont {Tommasini}, \citenamefont {Ali}, \citenamefont {Amendt}, \citenamefont {Atherton}, \citenamefont {Bachmann}, \citenamefont {Bailey}, \citenamefont {Benedetti}, \citenamefont {Berzak~Hopkins}, \citenamefont {Betti}, \citenamefont {Bhandarkar}, \citenamefont {Biener}, \citenamefont {Bionta}, \citenamefont {Birge}, \citenamefont {Bond}, \citenamefont {Bradley}, \citenamefont {Braun}, \citenamefont {Briggs}, \citenamefont {Bruhn}, \citenamefont {Celliers}, \citenamefont {Chang}, \citenamefont {Chapman}, \citenamefont {Chen}, \citenamefont {Choate},
  \citenamefont {Christopherson}, \citenamefont {Clark}, \citenamefont {Crippen}, \citenamefont {Dewald}, \citenamefont {Dittrich}, \citenamefont {Edwards}, \citenamefont {Farmer}, \citenamefont {Field}, \citenamefont {Fittinghoff}, \citenamefont {Frenje}, \citenamefont {Gaffney}, \citenamefont {Gatu~Johnson}, \citenamefont {Glenzer}, \citenamefont {Grim}, \citenamefont {Haan}, \citenamefont {Hahn}, \citenamefont {Hall}, \citenamefont {Hammel}, \citenamefont {Harte}, \citenamefont {Hartouni}, \citenamefont {Heebner}, \citenamefont {Hernandez}, \citenamefont {Herrmann}, \citenamefont {Herrmann}, \citenamefont {Hinkel}, \citenamefont {Ho}, \citenamefont {Holder}, \citenamefont {Hsing}, \citenamefont {Huang}, \citenamefont {Humbird}, \citenamefont {Izumi}, \citenamefont {Jarrott}, \citenamefont {Jeet}, \citenamefont {Jones}, \citenamefont {Kerbel}, \citenamefont {Kerr}, \citenamefont {Khan}, \citenamefont {Kilkenny}, \citenamefont {Kim}, \citenamefont {Geppert~Kleinrath}, \citenamefont {Geppert~Kleinrath},
  \citenamefont {Kong}, \citenamefont {Koning}, \citenamefont {Kroll}, \citenamefont {Kruse}, \citenamefont {Kustowski}, \citenamefont {Landen}, \citenamefont {Langer}, \citenamefont {Larson}, \citenamefont {Lemos}, \citenamefont {Lindl}, \citenamefont {Ma}, \citenamefont {MacDonald}, \citenamefont {MacGowan}, \citenamefont {Mackinnon}, \citenamefont {MacLaren}, \citenamefont {MacPhee}, \citenamefont {Marinak}, \citenamefont {Mariscal}, \citenamefont {Marley}, \citenamefont {Masse}, \citenamefont {Meaney}, \citenamefont {Meezan}, \citenamefont {Michel}, \citenamefont {Millot}, \citenamefont {Milovich}, \citenamefont {Moody}, \citenamefont {Moore}, \citenamefont {Morton}, \citenamefont {Murphy}, \citenamefont {Newman}, \citenamefont {Di~Nicola}, \citenamefont {Nikroo}, \citenamefont {Nora}, \citenamefont {Patel}, \citenamefont {Pelz}, \citenamefont {Peterson}, \citenamefont {Ping}, \citenamefont {Pollock}, \citenamefont {Ratledge}, \citenamefont {Rice}, \citenamefont {Rinderknecht}, \citenamefont {Rosen},
  \citenamefont {Rubery}, \citenamefont {Salmonson}, \citenamefont {Sater}, \citenamefont {Schiaffino}, \citenamefont {Schlossberg}, \citenamefont {Schneider}, \citenamefont {Schroeder}, \citenamefont {Scott}, \citenamefont {Sepke}, \citenamefont {Sequoia}, \citenamefont {Sherlock}, \citenamefont {Shin}, \citenamefont {Smalyuk}, \citenamefont {Spears}, \citenamefont {Springer}, \citenamefont {Stadermann}, \citenamefont {Stoupin}, \citenamefont {Strozzi}, \citenamefont {Suter}, \citenamefont {Thomas}, \citenamefont {Town}, \citenamefont {Tubman}, \citenamefont {Trosseille}, \citenamefont {Volegov}, \citenamefont {Weber}, \citenamefont {Widmann}, \citenamefont {Wild}, \citenamefont {Wilde}, \citenamefont {Van~Wonterghem}, \citenamefont {Woods}, \citenamefont {Woodworth}, \citenamefont {Yamaguchi}, \citenamefont {Yang},\ and\ \citenamefont {Zimmerman}}]{Zylstra2022}%
  \BibitemOpen
  \bibfield  {author} {\bibinfo {author} {\bibfnamefont {A.~B.}\ \bibnamefont {Zylstra}}, \bibinfo {author} {\bibfnamefont {O.~A.}\ \bibnamefont {Hurricane}}, \bibinfo {author} {\bibfnamefont {D.~A.}\ \bibnamefont {Callahan}}, \bibinfo {author} {\bibfnamefont {A.~L.}\ \bibnamefont {Kritcher}}, \bibinfo {author} {\bibfnamefont {J.~E.}\ \bibnamefont {Ralph}}, \bibinfo {author} {\bibfnamefont {H.~F.}\ \bibnamefont {Robey}}, \bibinfo {author} {\bibfnamefont {J.~S.}\ \bibnamefont {Ross}}, \bibinfo {author} {\bibfnamefont {C.~V.}\ \bibnamefont {Young}}, \bibinfo {author} {\bibfnamefont {K.~L.}\ \bibnamefont {Baker}}, \bibinfo {author} {\bibfnamefont {D.~T.}\ \bibnamefont {Casey}}, \bibinfo {author} {\bibfnamefont {T.}~\bibnamefont {D{\"o}ppner}}, \bibinfo {author} {\bibfnamefont {L.}~\bibnamefont {Divol}}, \bibinfo {author} {\bibfnamefont {M.}~\bibnamefont {Hohenberger}}, \bibinfo {author} {\bibfnamefont {S.}~\bibnamefont {Le~Pape}}, \bibinfo {author} {\bibfnamefont {A.}~\bibnamefont {Pak}}, \bibinfo {author}
  {\bibfnamefont {P.~K.}\ \bibnamefont {Patel}}, \bibinfo {author} {\bibfnamefont {R.}~\bibnamefont {Tommasini}}, \bibinfo {author} {\bibfnamefont {S.~J.}\ \bibnamefont {Ali}}, \bibinfo {author} {\bibfnamefont {P.~A.}\ \bibnamefont {Amendt}}, \bibinfo {author} {\bibfnamefont {L.~J.}\ \bibnamefont {Atherton}}, \bibinfo {author} {\bibfnamefont {B.}~\bibnamefont {Bachmann}}, \bibinfo {author} {\bibfnamefont {D.}~\bibnamefont {Bailey}}, \bibinfo {author} {\bibfnamefont {L.~R.}\ \bibnamefont {Benedetti}}, \bibinfo {author} {\bibfnamefont {L.}~\bibnamefont {Berzak~Hopkins}}, \bibinfo {author} {\bibfnamefont {R.}~\bibnamefont {Betti}}, \bibinfo {author} {\bibfnamefont {S.~D.}\ \bibnamefont {Bhandarkar}}, \bibinfo {author} {\bibfnamefont {J.}~\bibnamefont {Biener}}, \bibinfo {author} {\bibfnamefont {R.~M.}\ \bibnamefont {Bionta}}, \bibinfo {author} {\bibfnamefont {N.~W.}\ \bibnamefont {Birge}}, \bibinfo {author} {\bibfnamefont {E.~J.}\ \bibnamefont {Bond}}, \bibinfo {author} {\bibfnamefont {D.~K.}\ \bibnamefont
  {Bradley}}, \bibinfo {author} {\bibfnamefont {T.}~\bibnamefont {Braun}}, \bibinfo {author} {\bibfnamefont {T.~M.}\ \bibnamefont {Briggs}}, \bibinfo {author} {\bibfnamefont {M.~W.}\ \bibnamefont {Bruhn}}, \bibinfo {author} {\bibfnamefont {P.~M.}\ \bibnamefont {Celliers}}, \bibinfo {author} {\bibfnamefont {B.}~\bibnamefont {Chang}}, \bibinfo {author} {\bibfnamefont {T.}~\bibnamefont {Chapman}}, \bibinfo {author} {\bibfnamefont {H.}~\bibnamefont {Chen}}, \bibinfo {author} {\bibfnamefont {C.}~\bibnamefont {Choate}}, \bibinfo {author} {\bibfnamefont {A.~R.}\ \bibnamefont {Christopherson}}, \bibinfo {author} {\bibfnamefont {D.~S.}\ \bibnamefont {Clark}}, \bibinfo {author} {\bibfnamefont {J.~W.}\ \bibnamefont {Crippen}}, \bibinfo {author} {\bibfnamefont {E.~L.}\ \bibnamefont {Dewald}}, \bibinfo {author} {\bibfnamefont {T.~R.}\ \bibnamefont {Dittrich}}, \bibinfo {author} {\bibfnamefont {M.~J.}\ \bibnamefont {Edwards}}, \bibinfo {author} {\bibfnamefont {W.~A.}\ \bibnamefont {Farmer}}, \bibinfo {author}
  {\bibfnamefont {J.~E.}\ \bibnamefont {Field}}, \bibinfo {author} {\bibfnamefont {D.}~\bibnamefont {Fittinghoff}}, \bibinfo {author} {\bibfnamefont {J.}~\bibnamefont {Frenje}}, \bibinfo {author} {\bibfnamefont {J.}~\bibnamefont {Gaffney}}, \bibinfo {author} {\bibfnamefont {M.}~\bibnamefont {Gatu~Johnson}}, \bibinfo {author} {\bibfnamefont {S.~H.}\ \bibnamefont {Glenzer}}, \bibinfo {author} {\bibfnamefont {G.~P.}\ \bibnamefont {Grim}}, \bibinfo {author} {\bibfnamefont {S.}~\bibnamefont {Haan}}, \bibinfo {author} {\bibfnamefont {K.~D.}\ \bibnamefont {Hahn}}, \bibinfo {author} {\bibfnamefont {G.~N.}\ \bibnamefont {Hall}}, \bibinfo {author} {\bibfnamefont {B.~A.}\ \bibnamefont {Hammel}}, \bibinfo {author} {\bibfnamefont {J.}~\bibnamefont {Harte}}, \bibinfo {author} {\bibfnamefont {E.}~\bibnamefont {Hartouni}}, \bibinfo {author} {\bibfnamefont {J.~E.}\ \bibnamefont {Heebner}}, \bibinfo {author} {\bibfnamefont {V.~J.}\ \bibnamefont {Hernandez}}, \bibinfo {author} {\bibfnamefont {H.}~\bibnamefont {Herrmann}},
  \bibinfo {author} {\bibfnamefont {M.~C.}\ \bibnamefont {Herrmann}}, \bibinfo {author} {\bibfnamefont {D.~E.}\ \bibnamefont {Hinkel}}, \bibinfo {author} {\bibfnamefont {D.~D.}\ \bibnamefont {Ho}}, \bibinfo {author} {\bibfnamefont {J.~P.}\ \bibnamefont {Holder}}, \bibinfo {author} {\bibfnamefont {W.~W.}\ \bibnamefont {Hsing}}, \bibinfo {author} {\bibfnamefont {H.}~\bibnamefont {Huang}}, \bibinfo {author} {\bibfnamefont {K.~D.}\ \bibnamefont {Humbird}}, \bibinfo {author} {\bibfnamefont {N.}~\bibnamefont {Izumi}}, \bibinfo {author} {\bibfnamefont {L.~C.}\ \bibnamefont {Jarrott}}, \bibinfo {author} {\bibfnamefont {J.}~\bibnamefont {Jeet}}, \bibinfo {author} {\bibfnamefont {O.}~\bibnamefont {Jones}}, \bibinfo {author} {\bibfnamefont {G.~D.}\ \bibnamefont {Kerbel}}, \bibinfo {author} {\bibfnamefont {S.~M.}\ \bibnamefont {Kerr}}, \bibinfo {author} {\bibfnamefont {S.~F.}\ \bibnamefont {Khan}}, \bibinfo {author} {\bibfnamefont {J.}~\bibnamefont {Kilkenny}}, \bibinfo {author} {\bibfnamefont {Y.}~\bibnamefont {Kim}},
  \bibinfo {author} {\bibfnamefont {H.}~\bibnamefont {Geppert~Kleinrath}}, \bibinfo {author} {\bibfnamefont {V.}~\bibnamefont {Geppert~Kleinrath}}, \bibinfo {author} {\bibfnamefont {C.}~\bibnamefont {Kong}}, \bibinfo {author} {\bibfnamefont {J.~M.}\ \bibnamefont {Koning}}, \bibinfo {author} {\bibfnamefont {J.~J.}\ \bibnamefont {Kroll}}, \bibinfo {author} {\bibfnamefont {M.~K.~G.}\ \bibnamefont {Kruse}}, \bibinfo {author} {\bibfnamefont {B.}~\bibnamefont {Kustowski}}, \bibinfo {author} {\bibfnamefont {O.~L.}\ \bibnamefont {Landen}}, \bibinfo {author} {\bibfnamefont {S.}~\bibnamefont {Langer}}, \bibinfo {author} {\bibfnamefont {D.}~\bibnamefont {Larson}}, \bibinfo {author} {\bibfnamefont {N.~C.}\ \bibnamefont {Lemos}}, \bibinfo {author} {\bibfnamefont {J.~D.}\ \bibnamefont {Lindl}}, \bibinfo {author} {\bibfnamefont {T.}~\bibnamefont {Ma}}, \bibinfo {author} {\bibfnamefont {M.~J.}\ \bibnamefont {MacDonald}}, \bibinfo {author} {\bibfnamefont {B.~J.}\ \bibnamefont {MacGowan}}, \bibinfo {author} {\bibfnamefont
  {A.~J.}\ \bibnamefont {Mackinnon}}, \bibinfo {author} {\bibfnamefont {S.~A.}\ \bibnamefont {MacLaren}}, \bibinfo {author} {\bibfnamefont {A.~G.}\ \bibnamefont {MacPhee}}, \bibinfo {author} {\bibfnamefont {M.~M.}\ \bibnamefont {Marinak}}, \bibinfo {author} {\bibfnamefont {D.~A.}\ \bibnamefont {Mariscal}}, \bibinfo {author} {\bibfnamefont {E.~V.}\ \bibnamefont {Marley}}, \bibinfo {author} {\bibfnamefont {L.}~\bibnamefont {Masse}}, \bibinfo {author} {\bibfnamefont {K.}~\bibnamefont {Meaney}}, \bibinfo {author} {\bibfnamefont {N.~B.}\ \bibnamefont {Meezan}}, \bibinfo {author} {\bibfnamefont {P.~A.}\ \bibnamefont {Michel}}, \bibinfo {author} {\bibfnamefont {M.}~\bibnamefont {Millot}}, \bibinfo {author} {\bibfnamefont {J.~L.}\ \bibnamefont {Milovich}}, \bibinfo {author} {\bibfnamefont {J.~D.}\ \bibnamefont {Moody}}, \bibinfo {author} {\bibfnamefont {A.~S.}\ \bibnamefont {Moore}}, \bibinfo {author} {\bibfnamefont {J.~W.}\ \bibnamefont {Morton}}, \bibinfo {author} {\bibfnamefont {T.}~\bibnamefont {Murphy}},
  \bibinfo {author} {\bibfnamefont {K.}~\bibnamefont {Newman}}, \bibinfo {author} {\bibfnamefont {J.-M.~G.}\ \bibnamefont {Di~Nicola}}, \bibinfo {author} {\bibfnamefont {A.}~\bibnamefont {Nikroo}}, \bibinfo {author} {\bibfnamefont {R.}~\bibnamefont {Nora}}, \bibinfo {author} {\bibfnamefont {M.~V.}\ \bibnamefont {Patel}}, \bibinfo {author} {\bibfnamefont {L.~J.}\ \bibnamefont {Pelz}}, \bibinfo {author} {\bibfnamefont {J.~L.}\ \bibnamefont {Peterson}}, \bibinfo {author} {\bibfnamefont {Y.}~\bibnamefont {Ping}}, \bibinfo {author} {\bibfnamefont {B.~B.}\ \bibnamefont {Pollock}}, \bibinfo {author} {\bibfnamefont {M.}~\bibnamefont {Ratledge}}, \bibinfo {author} {\bibfnamefont {N.~G.}\ \bibnamefont {Rice}}, \bibinfo {author} {\bibfnamefont {H.}~\bibnamefont {Rinderknecht}}, \bibinfo {author} {\bibfnamefont {M.}~\bibnamefont {Rosen}}, \bibinfo {author} {\bibfnamefont {M.~S.}\ \bibnamefont {Rubery}}, \bibinfo {author} {\bibfnamefont {J.~D.}\ \bibnamefont {Salmonson}}, \bibinfo {author} {\bibfnamefont {J.}~\bibnamefont
  {Sater}}, \bibinfo {author} {\bibfnamefont {S.}~\bibnamefont {Schiaffino}}, \bibinfo {author} {\bibfnamefont {D.~J.}\ \bibnamefont {Schlossberg}}, \bibinfo {author} {\bibfnamefont {M.~B.}\ \bibnamefont {Schneider}}, \bibinfo {author} {\bibfnamefont {C.~R.}\ \bibnamefont {Schroeder}}, \bibinfo {author} {\bibfnamefont {H.~A.}\ \bibnamefont {Scott}}, \bibinfo {author} {\bibfnamefont {S.~M.}\ \bibnamefont {Sepke}}, \bibinfo {author} {\bibfnamefont {K.}~\bibnamefont {Sequoia}}, \bibinfo {author} {\bibfnamefont {M.~W.}\ \bibnamefont {Sherlock}}, \bibinfo {author} {\bibfnamefont {S.}~\bibnamefont {Shin}}, \bibinfo {author} {\bibfnamefont {V.~A.}\ \bibnamefont {Smalyuk}}, \bibinfo {author} {\bibfnamefont {B.~K.}\ \bibnamefont {Spears}}, \bibinfo {author} {\bibfnamefont {P.~T.}\ \bibnamefont {Springer}}, \bibinfo {author} {\bibfnamefont {M.}~\bibnamefont {Stadermann}}, \bibinfo {author} {\bibfnamefont {S.}~\bibnamefont {Stoupin}}, \bibinfo {author} {\bibfnamefont {D.~J.}\ \bibnamefont {Strozzi}}, \bibinfo {author}
  {\bibfnamefont {L.~J.}\ \bibnamefont {Suter}}, \bibinfo {author} {\bibfnamefont {C.~A.}\ \bibnamefont {Thomas}}, \bibinfo {author} {\bibfnamefont {R.~P.~J.}\ \bibnamefont {Town}}, \bibinfo {author} {\bibfnamefont {E.~R.}\ \bibnamefont {Tubman}}, \bibinfo {author} {\bibfnamefont {C.}~\bibnamefont {Trosseille}}, \bibinfo {author} {\bibfnamefont {P.~L.}\ \bibnamefont {Volegov}}, \bibinfo {author} {\bibfnamefont {C.~R.}\ \bibnamefont {Weber}}, \bibinfo {author} {\bibfnamefont {K.}~\bibnamefont {Widmann}}, \bibinfo {author} {\bibfnamefont {C.}~\bibnamefont {Wild}}, \bibinfo {author} {\bibfnamefont {C.~H.}\ \bibnamefont {Wilde}}, \bibinfo {author} {\bibfnamefont {B.~M.}\ \bibnamefont {Van~Wonterghem}}, \bibinfo {author} {\bibfnamefont {D.~T.}\ \bibnamefont {Woods}}, \bibinfo {author} {\bibfnamefont {B.~N.}\ \bibnamefont {Woodworth}}, \bibinfo {author} {\bibfnamefont {M.}~\bibnamefont {Yamaguchi}}, \bibinfo {author} {\bibfnamefont {S.~T.}\ \bibnamefont {Yang}},\ and\ \bibinfo {author} {\bibfnamefont {G.~B.}\
  \bibnamefont {Zimmerman}},\ }\bibfield  {title} {\enquote {\bibinfo {title} {Burning plasma achieved in inertial fusion},}\ }\href {https://doi.org/10.1038/s41586-021-04281-w} {\bibfield  {journal} {\bibinfo  {journal} {Nature}\ }\textbf {\bibinfo {volume} {601}},\ \bibinfo {pages} {542--548} (\bibinfo {year} {2022})}\BibitemShut {NoStop}%
\bibitem [{\citenamefont {Qiao}\ and\ \citenamefont {Lan}(2021)}]{PhysRevLett.126.185001}%
  \BibitemOpen
  \bibfield  {author} {\bibinfo {author} {\bibfnamefont {X.}~\bibnamefont {Qiao}}\ and\ \bibinfo {author} {\bibfnamefont {K.}~\bibnamefont {Lan}},\ }\bibfield  {title} {\enquote {\bibinfo {title} {Novel target designs to mitigate hydrodynamic instabilities growth in inertial confinement fusion},}\ }\href {https://doi.org/10.1103/PhysRevLett.126.185001} {\bibfield  {journal} {\bibinfo  {journal} {Phys. Rev. Lett.}\ }\textbf {\bibinfo {volume} {126}},\ \bibinfo {pages} {185001} (\bibinfo {year} {2021})}\BibitemShut {NoStop}%
\bibitem [{\citenamefont {Yan}\ \emph {et~al.}(2012)\citenamefont {Yan}, \citenamefont {Jiang}, \citenamefont {Su}, \citenamefont {Wu},\ and\ \citenamefont {Lin}}]{2012-6-068703}%
  \BibitemOpen
  \bibfield  {author} {\bibinfo {author} {\bibfnamefont {J.}~\bibnamefont {Yan}}, \bibinfo {author} {\bibfnamefont {S.-E.}\ \bibnamefont {Jiang}}, \bibinfo {author} {\bibfnamefont {M.}~\bibnamefont {Su}}, \bibinfo {author} {\bibfnamefont {S.-C.}\ \bibnamefont {Wu}},\ and\ \bibinfo {author} {\bibfnamefont {Z.-W.}\ \bibnamefont {Lin}},\ }\bibfield  {title} {\enquote {\bibinfo {title} {The application of phase contrast imaging to icf multi-shell capsule diagnosis},}\ }\href {https://doi.org/10.7498/aps.61.068703} {\bibfield  {journal} {\bibinfo  {journal} {Acta Physica Sinica}\ }\textbf {\bibinfo {volume} {61}},\ \bibinfo {pages} {068703} (\bibinfo {year} {2012})}\BibitemShut {NoStop}%
\bibitem [{\citenamefont {Wang}\ \emph {et~al.}(2019)\citenamefont {Wang}, \citenamefont {Wang}, \citenamefont {Ma},\ and\ \citenamefont {Zhao}}]{10.1063/1.5085863}%
  \BibitemOpen
  \bibfield  {author} {\bibinfo {author} {\bibfnamefont {L.}~\bibnamefont {Wang}}, \bibinfo {author} {\bibfnamefont {Y.}~\bibnamefont {Wang}}, \bibinfo {author} {\bibfnamefont {X.}~\bibnamefont {Ma}},\ and\ \bibinfo {author} {\bibfnamefont {W.}~\bibnamefont {Zhao}},\ }\bibfield  {title} {\enquote {\bibinfo {title} {{Measurement of laser differential confocal geometrical parameters for ICF capsule}},}\ }\href {https://doi.org/10.1063/1.5085863} {\bibfield  {journal} {\bibinfo  {journal} {Matter and Radiation at Extremes}\ }\textbf {\bibinfo {volume} {4}},\ \bibinfo {pages} {025401} (\bibinfo {year} {2019})},\ \Eprint {https://arxiv.org/abs/https://pubs.aip.org/aip/mre/article-pdf/doi/10.1063/1.5085863/13909652/025401\_1\_online.pdf} {https://pubs.aip.org/aip/mre/article-pdf/doi/10.1063/1.5085863/13909652/025401\_1\_online.pdf} \BibitemShut {NoStop}%
\bibitem [{\citenamefont {Feng}\ \emph {et~al.}(2020)\citenamefont {Feng}, \citenamefont {Xing}, \citenamefont {Yulong}, \citenamefont {Bolun}, \citenamefont {Zhongjing}, \citenamefont {Tao}, \citenamefont {Xincheng}, \citenamefont {Hang}, \citenamefont {Kuan}, \citenamefont {Jiamin}, \citenamefont {Shaoen},\ and\ \citenamefont {Baohan}}]{10.11884/HPLPB202032.200136}%
  \BibitemOpen
  \bibfield  {author} {\bibinfo {author} {\bibfnamefont {W.}~\bibnamefont {Feng}}, \bibinfo {author} {\bibfnamefont {Z.}~\bibnamefont {Xing}}, \bibinfo {author} {\bibfnamefont {L.}~\bibnamefont {Yulong}}, \bibinfo {author} {\bibfnamefont {C.}~\bibnamefont {Bolun}}, \bibinfo {author} {\bibfnamefont {C.}~\bibnamefont {Zhongjing}}, \bibinfo {author} {\bibfnamefont {X.}~\bibnamefont {Tao}}, \bibinfo {author} {\bibfnamefont {L.}~\bibnamefont {Xincheng}}, \bibinfo {author} {\bibfnamefont {Z.}~\bibnamefont {Hang}}, \bibinfo {author} {\bibfnamefont {R.}~\bibnamefont {Kuan}}, \bibinfo {author} {\bibfnamefont {Y.}~\bibnamefont {Jiamin}}, \bibinfo {author} {\bibfnamefont {J.}~\bibnamefont {Shaoen}},\ and\ \bibinfo {author} {\bibfnamefont {Z.}~\bibnamefont {Baohan}},\ }\bibfield  {title} {\enquote {\bibinfo {title} {{Progress in high time- and space-resolving diagnostic technique for laser-driven inertial confinement fusion}},}\ }\href {https://doi.org/10.11884/HPLPB202032.200136} {\bibfield  {journal} {\bibinfo
  {journal} {High Power laser and Particle Beams}\ }\textbf {\bibinfo {volume} {32}},\ \bibinfo {pages} {112002} (\bibinfo {year} {2020})}\BibitemShut {NoStop}%
\bibitem [{\citenamefont {Jing-Yu}, \citenamefont {Lu-Ping},\ and\ \citenamefont {Yang}(2020)}]{2020-49-10-001}%
  \BibitemOpen
  \bibfield  {author} {\bibinfo {author} {\bibfnamefont {T.}~\bibnamefont {Jing-Yu}}, \bibinfo {author} {\bibfnamefont {Z.}~\bibnamefont {Lu-Ping}},\ and\ \bibinfo {author} {\bibfnamefont {H.}~\bibnamefont {Yang}},\ }\bibfield  {title} {\enquote {\bibinfo {title} {Multidisciplinary research and applications of muon sources},}\ }\href {https://doi.org/10.7693/wl20201001} {\bibfield  {journal} {\bibinfo  {journal} {Physics}\ }\textbf {\bibinfo {volume} {49}},\ \bibinfo {pages} {645--656} (\bibinfo {year} {2020})}\BibitemShut {NoStop}%
\bibitem [{\citenamefont {Jian}, \citenamefont {Liang},\ and\ \citenamefont {Ye}(2021)}]{2021-50-4-006}%
  \BibitemOpen
  \bibfield  {author} {\bibinfo {author} {\bibfnamefont {T.}~\bibnamefont {Jian}}, \bibinfo {author} {\bibfnamefont {L.}~\bibnamefont {Liang}},\ and\ \bibinfo {author} {\bibfnamefont {Y.}~\bibnamefont {Ye}},\ }\bibfield  {title} {\enquote {\bibinfo {title} {Research status and prospect of muon physics experiments},}\ }\href {https://doi.org/10.7693/wl20210406} {\bibfield  {journal} {\bibinfo  {journal} {Physics}\ }\textbf {\bibinfo {volume} {50}},\ \bibinfo {pages} {239--247} (\bibinfo {year} {2021})}\BibitemShut {NoStop}%
\bibitem [{\citenamefont {Bang-Jiao}, \citenamefont {Yang},\ and\ \citenamefont {Zhi-Hao}(2021)}]{2021-50-4-007}%
  \BibitemOpen
  \bibfield  {author} {\bibinfo {author} {\bibfnamefont {Y.}~\bibnamefont {Bang-Jiao}}, \bibinfo {author} {\bibfnamefont {L.}~\bibnamefont {Yang}},\ and\ \bibinfo {author} {\bibfnamefont {Z.}~\bibnamefont {Zhi-Hao}},\ }\bibfield  {title} {\enquote {\bibinfo {title} {Muon imaging and elemental analysis},}\ }\href {https://doi.org/10.7693/wl20210407} {\bibfield  {journal} {\bibinfo  {journal} {Physics}\ }\textbf {\bibinfo {volume} {50}},\ \bibinfo {pages} {248--256} (\bibinfo {year} {2021})}\BibitemShut {NoStop}%
\bibitem [{\citenamefont {Lei}, \citenamefont {Xiao-Jie},\ and\ \citenamefont {Zi-Wen}(2021)}]{2021-50-4-008}%
  \BibitemOpen
  \bibfield  {author} {\bibinfo {author} {\bibfnamefont {S.}~\bibnamefont {Lei}}, \bibinfo {author} {\bibfnamefont {N.}~\bibnamefont {Xiao-Jie}},\ and\ \bibinfo {author} {\bibfnamefont {P.}~\bibnamefont {Zi-Wen}},\ }\bibfield  {title} {\enquote {\bibinfo {title} {Application of muon spin relaxation/rotation in condensed matter physics},}\ }\href {https://doi.org/10.7693/wl20210408} {\bibfield  {journal} {\bibinfo  {journal} {Physics}\ }\textbf {\bibinfo {volume} {50}},\ \bibinfo {pages} {257--265} (\bibinfo {year} {2021})}\BibitemShut {NoStop}%
\bibitem [{\citenamefont {Chang}(1949)}]{RevModPhys.21.166}%
  \BibitemOpen
  \bibfield  {author} {\bibinfo {author} {\bibfnamefont {W.~Y.}\ \bibnamefont {Chang}},\ }\bibfield  {title} {\enquote {\bibinfo {title} {A cloud-chamber study of meson absorption by thin pb, fe, and al foils},}\ }\href {https://doi.org/10.1103/RevModPhys.21.166} {\bibfield  {journal} {\bibinfo  {journal} {Rev. Mod. Phys.}\ }\textbf {\bibinfo {volume} {21}},\ \bibinfo {pages} {166--180} (\bibinfo {year} {1949})}\BibitemShut {NoStop}%
\bibitem [{\citenamefont {Hillier}, \citenamefont {Paul},\ and\ \citenamefont {Ishida}(2016)}]{HILLIER2016203}%
  \BibitemOpen
  \bibfield  {author} {\bibinfo {author} {\bibfnamefont {A.}~\bibnamefont {Hillier}}, \bibinfo {author} {\bibfnamefont {D.}~\bibnamefont {Paul}},\ and\ \bibinfo {author} {\bibfnamefont {K.}~\bibnamefont {Ishida}},\ }\bibfield  {title} {\enquote {\bibinfo {title} {Probing beneath the surface without a scratch — bulk non-destructive elemental analysis using negative muons},}\ }\href {https://doi.org/https://doi.org/10.1016/j.microc.2015.11.031} {\bibfield  {journal} {\bibinfo  {journal} {Microchemical Journal}\ }\textbf {\bibinfo {volume} {125}},\ \bibinfo {pages} {203--207} (\bibinfo {year} {2016})}\BibitemShut {NoStop}%
\bibitem [{\citenamefont {Ninomiya}\ \emph {et~al.}(2010)\citenamefont {Ninomiya}, \citenamefont {Nagatomo}, \citenamefont {Kubo}, \citenamefont {Strasser}, \citenamefont {Kawamura}, \citenamefont {Shimomura}, \citenamefont {Miyake}, \citenamefont {Saito},\ and\ \citenamefont {Higemoto}}]{Ninomiya_2010}%
  \BibitemOpen
  \bibfield  {author} {\bibinfo {author} {\bibfnamefont {K.}~\bibnamefont {Ninomiya}}, \bibinfo {author} {\bibfnamefont {T.}~\bibnamefont {Nagatomo}}, \bibinfo {author} {\bibfnamefont {K.~M.}\ \bibnamefont {Kubo}}, \bibinfo {author} {\bibfnamefont {P.}~\bibnamefont {Strasser}}, \bibinfo {author} {\bibfnamefont {N.}~\bibnamefont {Kawamura}}, \bibinfo {author} {\bibfnamefont {K.}~\bibnamefont {Shimomura}}, \bibinfo {author} {\bibfnamefont {Y.}~\bibnamefont {Miyake}}, \bibinfo {author} {\bibfnamefont {T.}~\bibnamefont {Saito}},\ and\ \bibinfo {author} {\bibfnamefont {W.}~\bibnamefont {Higemoto}},\ }\bibfield  {title} {\enquote {\bibinfo {title} {Development of elemental analysis by muonic x-ray measurement in j-parc},}\ }\href {https://doi.org/10.1088/1742-6596/225/1/012040} {\bibfield  {journal} {\bibinfo  {journal} {Journal of Physics: Conference Series}\ }\textbf {\bibinfo {volume} {225}},\ \bibinfo {pages} {012040} (\bibinfo {year} {2010})}\BibitemShut {NoStop}%
\bibitem [{\citenamefont {Clemenza}\ \emph {et~al.}(2019)\citenamefont {Clemenza}, \citenamefont {Bonesini}, \citenamefont {Carpinelli}, \citenamefont {Cremonesi}, \citenamefont {Fiorini}, \citenamefont {Gorini}, \citenamefont {Hillier}, \citenamefont {Ishida}, \citenamefont {Menegolli}, \citenamefont {Mocchiutti}, \citenamefont {Oliva}, \citenamefont {Prata}, \citenamefont {Rendeli}, \citenamefont {Rignanese}, \citenamefont {Rossella}, \citenamefont {Sipala}, \citenamefont {Soldani}, \citenamefont {Tortora}, \citenamefont {Vacchi},\ and\ \citenamefont {Vallazza}}]{Clemenza2019}%
  \BibitemOpen
  \bibfield  {author} {\bibinfo {author} {\bibfnamefont {M.}~\bibnamefont {Clemenza}}, \bibinfo {author} {\bibfnamefont {M.}~\bibnamefont {Bonesini}}, \bibinfo {author} {\bibfnamefont {M.}~\bibnamefont {Carpinelli}}, \bibinfo {author} {\bibfnamefont {O.}~\bibnamefont {Cremonesi}}, \bibinfo {author} {\bibfnamefont {E.}~\bibnamefont {Fiorini}}, \bibinfo {author} {\bibfnamefont {G.}~\bibnamefont {Gorini}}, \bibinfo {author} {\bibfnamefont {A.}~\bibnamefont {Hillier}}, \bibinfo {author} {\bibfnamefont {K.}~\bibnamefont {Ishida}}, \bibinfo {author} {\bibfnamefont {A.}~\bibnamefont {Menegolli}}, \bibinfo {author} {\bibfnamefont {E.}~\bibnamefont {Mocchiutti}}, \bibinfo {author} {\bibfnamefont {P.}~\bibnamefont {Oliva}}, \bibinfo {author} {\bibfnamefont {M.}~\bibnamefont {Prata}}, \bibinfo {author} {\bibfnamefont {M.}~\bibnamefont {Rendeli}}, \bibinfo {author} {\bibfnamefont {L.~P.}\ \bibnamefont {Rignanese}}, \bibinfo {author} {\bibfnamefont {M.}~\bibnamefont {Rossella}}, \bibinfo {author} {\bibfnamefont
  {V.}~\bibnamefont {Sipala}}, \bibinfo {author} {\bibfnamefont {M.}~\bibnamefont {Soldani}}, \bibinfo {author} {\bibfnamefont {L.}~\bibnamefont {Tortora}}, \bibinfo {author} {\bibfnamefont {A.}~\bibnamefont {Vacchi}},\ and\ \bibinfo {author} {\bibfnamefont {E.}~\bibnamefont {Vallazza}},\ }\bibfield  {title} {\enquote {\bibinfo {title} {Muonic atom x-ray spectroscopy for non-destructive analysis of archeological samples},}\ }\href {https://doi.org/10.1007/s10967-019-06927-6} {\bibfield  {journal} {\bibinfo  {journal} {Journal of Radioanalytical and Nuclear Chemistry}\ }\textbf {\bibinfo {volume} {322}},\ \bibinfo {pages} {1357--1363} (\bibinfo {year} {2019})}\BibitemShut {NoStop}%
\bibitem [{\citenamefont {Daniel}(1987)}]{Daniel1987}%
  \BibitemOpen
  \bibfield  {author} {\bibinfo {author} {\bibfnamefont {H.}~\bibnamefont {Daniel}},\ }\bibfield  {title} {\enquote {\bibinfo {title} {Application of muonic x-rays for elemental analysis},}\ }\href {https://doi.org/10.1007/BF02796642} {\bibfield  {journal} {\bibinfo  {journal} {Biological Trace Element Research}\ }\textbf {\bibinfo {volume} {13}},\ \bibinfo {pages} {301--318} (\bibinfo {year} {1987})}\BibitemShut {NoStop}%
\bibitem [{\citenamefont {Terada}\ \emph {et~al.}(2017)\citenamefont {Terada}, \citenamefont {Sato}, \citenamefont {Ninomiya}, \citenamefont {Kawashima}, \citenamefont {Shimomura}, \citenamefont {Yoshida}, \citenamefont {Kawai}, \citenamefont {Osawa},\ and\ \citenamefont {Tachibana}}]{Terada2017}%
  \BibitemOpen
  \bibfield  {author} {\bibinfo {author} {\bibfnamefont {K.}~\bibnamefont {Terada}}, \bibinfo {author} {\bibfnamefont {A.}~\bibnamefont {Sato}}, \bibinfo {author} {\bibfnamefont {K.}~\bibnamefont {Ninomiya}}, \bibinfo {author} {\bibfnamefont {Y.}~\bibnamefont {Kawashima}}, \bibinfo {author} {\bibfnamefont {K.}~\bibnamefont {Shimomura}}, \bibinfo {author} {\bibfnamefont {G.}~\bibnamefont {Yoshida}}, \bibinfo {author} {\bibfnamefont {Y.}~\bibnamefont {Kawai}}, \bibinfo {author} {\bibfnamefont {T.}~\bibnamefont {Osawa}},\ and\ \bibinfo {author} {\bibfnamefont {S.}~\bibnamefont {Tachibana}},\ }\bibfield  {title} {\enquote {\bibinfo {title} {Non-destructive elemental analysis of a carbonaceous chondrite with direct current muon beam at music},}\ }\href {https://doi.org/10.1038/s41598-017-15719-5} {\bibfield  {journal} {\bibinfo  {journal} {Scientific Reports}\ }\textbf {\bibinfo {volume} {7}},\ \bibinfo {pages} {15478} (\bibinfo {year} {2017})}\BibitemShut {NoStop}%
\bibitem [{\citenamefont {Umegaki}\ \emph {et~al.}(2020)\citenamefont {Umegaki}, \citenamefont {Higuchi}, \citenamefont {Kondo}, \citenamefont {Ninomiya}, \citenamefont {Takeshita}, \citenamefont {Tampo}, \citenamefont {Nakano}, \citenamefont {Oka}, \citenamefont {Sugiyama}, \citenamefont {Kubo},\ and\ \citenamefont {Miyake}}]{doi:10.1021/acs.analchem.0c00370}%
  \BibitemOpen
  \bibfield  {author} {\bibinfo {author} {\bibfnamefont {I.}~\bibnamefont {Umegaki}}, \bibinfo {author} {\bibfnamefont {Y.}~\bibnamefont {Higuchi}}, \bibinfo {author} {\bibfnamefont {Y.}~\bibnamefont {Kondo}}, \bibinfo {author} {\bibfnamefont {K.}~\bibnamefont {Ninomiya}}, \bibinfo {author} {\bibfnamefont {S.}~\bibnamefont {Takeshita}}, \bibinfo {author} {\bibfnamefont {M.}~\bibnamefont {Tampo}}, \bibinfo {author} {\bibfnamefont {H.}~\bibnamefont {Nakano}}, \bibinfo {author} {\bibfnamefont {H.}~\bibnamefont {Oka}}, \bibinfo {author} {\bibfnamefont {J.}~\bibnamefont {Sugiyama}}, \bibinfo {author} {\bibfnamefont {M.~K.}\ \bibnamefont {Kubo}},\ and\ \bibinfo {author} {\bibfnamefont {Y.}~\bibnamefont {Miyake}},\ }\bibfield  {title} {\enquote {\bibinfo {title} {Nondestructive high-sensitivity detections of metallic lithium deposited on a battery anode using muonic x-rays},}\ }\href {https://doi.org/10.1021/acs.analchem.0c00370} {\bibfield  {journal} {\bibinfo  {journal} {Analytical Chemistry}\ }\textbf {\bibinfo
  {volume} {92}},\ \bibinfo {pages} {8194--8200} (\bibinfo {year} {2020})},\ \bibinfo {note} {pMID: 32468821},\ \Eprint {https://arxiv.org/abs/https://doi.org/10.1021/acs.analchem.0c00370} {https://doi.org/10.1021/acs.analchem.0c00370} \BibitemShut {NoStop}%
\bibitem [{\citenamefont {Aramini}\ \emph {et~al.}(2020)\citenamefont {Aramini}, \citenamefont {Milanese}, \citenamefont {Hillier}, \citenamefont {Girella}, \citenamefont {Horstmann}, \citenamefont {Klassen}, \citenamefont {Ishida}, \citenamefont {Dornheim},\ and\ \citenamefont {Pistidda}}]{nano10071260}%
  \BibitemOpen
  \bibfield  {author} {\bibinfo {author} {\bibfnamefont {M.}~\bibnamefont {Aramini}}, \bibinfo {author} {\bibfnamefont {C.}~\bibnamefont {Milanese}}, \bibinfo {author} {\bibfnamefont {A.~D.}\ \bibnamefont {Hillier}}, \bibinfo {author} {\bibfnamefont {A.}~\bibnamefont {Girella}}, \bibinfo {author} {\bibfnamefont {C.}~\bibnamefont {Horstmann}}, \bibinfo {author} {\bibfnamefont {T.}~\bibnamefont {Klassen}}, \bibinfo {author} {\bibfnamefont {K.}~\bibnamefont {Ishida}}, \bibinfo {author} {\bibfnamefont {M.}~\bibnamefont {Dornheim}},\ and\ \bibinfo {author} {\bibfnamefont {C.}~\bibnamefont {Pistidda}},\ }\bibfield  {title} {\enquote {\bibinfo {title} {Using the emission of muonic x-rays as a spectroscopic tool for the investigation of the local chemistry of elements},}\ }\href {https://doi.org/10.3390/nano10071260} {\bibfield  {journal} {\bibinfo  {journal} {Nanomaterials}\ }\textbf {\bibinfo {volume} {10}} (\bibinfo {year} {2020}),\ 10.3390/nano10071260}\BibitemShut {NoStop}%
\bibitem [{\citenamefont {Shimada-Takaura}\ \emph {et~al.}(2021)\citenamefont {Shimada-Takaura}, \citenamefont {Ninomiya}, \citenamefont {Sato}, \citenamefont {Ueda}, \citenamefont {Tampo}, \citenamefont {Takeshita}, \citenamefont {Umegaki}, \citenamefont {Miyake},\ and\ \citenamefont {Takahashi}}]{Shimada-Takaura2021}%
  \BibitemOpen
  \bibfield  {author} {\bibinfo {author} {\bibfnamefont {K.}~\bibnamefont {Shimada-Takaura}}, \bibinfo {author} {\bibfnamefont {K.}~\bibnamefont {Ninomiya}}, \bibinfo {author} {\bibfnamefont {A.}~\bibnamefont {Sato}}, \bibinfo {author} {\bibfnamefont {N.}~\bibnamefont {Ueda}}, \bibinfo {author} {\bibfnamefont {M.}~\bibnamefont {Tampo}}, \bibinfo {author} {\bibfnamefont {S.}~\bibnamefont {Takeshita}}, \bibinfo {author} {\bibfnamefont {I.}~\bibnamefont {Umegaki}}, \bibinfo {author} {\bibfnamefont {Y.}~\bibnamefont {Miyake}},\ and\ \bibinfo {author} {\bibfnamefont {K.}~\bibnamefont {Takahashi}},\ }\bibfield  {title} {\enquote {\bibinfo {title} {A novel challenge of nondestructive analysis on ogata koan's sealed medicine by muonic x-ray analysis},}\ }\href {https://doi.org/10.1007/s11418-021-01487-0} {\bibfield  {journal} {\bibinfo  {journal} {Journal of Natural Medicines}\ }\textbf {\bibinfo {volume} {75}},\ \bibinfo {pages} {532--539} (\bibinfo {year} {2021})}\BibitemShut {NoStop}%
\bibitem [{\citenamefont {Reidy}\ \emph {et~al.}(1978)\citenamefont {Reidy}, \citenamefont {Hutson}, \citenamefont {Daniel},\ and\ \citenamefont {Springer}}]{doi:10.1021/ac50023a015}%
  \BibitemOpen
  \bibfield  {author} {\bibinfo {author} {\bibfnamefont {J.~J.}\ \bibnamefont {Reidy}}, \bibinfo {author} {\bibfnamefont {R.~L.}\ \bibnamefont {Hutson}}, \bibinfo {author} {\bibfnamefont {H.}~\bibnamefont {Daniel}},\ and\ \bibinfo {author} {\bibfnamefont {K.}~\bibnamefont {Springer}},\ }\bibfield  {title} {\enquote {\bibinfo {title} {Use of muonic x-rays for nondestructive analysis of bulk samples for low z constituents},}\ }\href {https://doi.org/10.1021/ac50023a015} {\bibfield  {journal} {\bibinfo  {journal} {Analytical Chemistry}\ }\textbf {\bibinfo {volume} {50}},\ \bibinfo {pages} {40--44} (\bibinfo {year} {1978})},\ \Eprint {https://arxiv.org/abs/https://doi.org/10.1021/ac50023a015} {https://doi.org/10.1021/ac50023a015} \BibitemShut {NoStop}%
\bibitem [{\citenamefont {Hosoi}\ \emph {et~al.}(2014)\citenamefont {Hosoi}, \citenamefont {Watanabe}, \citenamefont {Sugita}, \citenamefont {Tanaka}, \citenamefont {Nagamine}, \citenamefont {Ono},\ and\ \citenamefont {Sakamoto}}]{10.1259/0007-1285-68-816-1325}%
  \BibitemOpen
  \bibfield  {author} {\bibinfo {author} {\bibfnamefont {Y.}~\bibnamefont {Hosoi}}, \bibinfo {author} {\bibfnamefont {Y.}~\bibnamefont {Watanabe}}, \bibinfo {author} {\bibfnamefont {R.}~\bibnamefont {Sugita}}, \bibinfo {author} {\bibfnamefont {Y.}~\bibnamefont {Tanaka}}, \bibinfo {author} {\bibfnamefont {K.}~\bibnamefont {Nagamine}}, \bibinfo {author} {\bibfnamefont {T.}~\bibnamefont {Ono}},\ and\ \bibinfo {author} {\bibfnamefont {K.}~\bibnamefont {Sakamoto}},\ }\bibfield  {title} {\enquote {\bibinfo {title} {{Non-destructive elemental analysis of vertebral body trabecular bone using muonic X-rays}},}\ }\href {https://doi.org/10.1259/0007-1285-68-816-1325} {\bibfield  {journal} {\bibinfo  {journal} {British Journal of Radiology}\ }\textbf {\bibinfo {volume} {68}},\ \bibinfo {pages} {1325--1331} (\bibinfo {year} {2014})},\ \Eprint {https://arxiv.org/abs/https://academic.oup.com/bjr/article-pdf/68/816/1325/55003839/0007-1285-68-816-1325.pdf}
  {https://academic.oup.com/bjr/article-pdf/68/816/1325/55003839/0007-1285-68-816-1325.pdf} \BibitemShut {NoStop}%
\bibitem [{\citenamefont {Hampshire}\ \emph {et~al.}(2019)\citenamefont {Hampshire}, \citenamefont {Butcher}, \citenamefont {Ishida}, \citenamefont {Green}, \citenamefont {Paul},\ and\ \citenamefont {Hillier}}]{heritage2010028}%
  \BibitemOpen
  \bibfield  {author} {\bibinfo {author} {\bibfnamefont {B.~V.}\ \bibnamefont {Hampshire}}, \bibinfo {author} {\bibfnamefont {K.}~\bibnamefont {Butcher}}, \bibinfo {author} {\bibfnamefont {K.}~\bibnamefont {Ishida}}, \bibinfo {author} {\bibfnamefont {G.}~\bibnamefont {Green}}, \bibinfo {author} {\bibfnamefont {D.~M.}\ \bibnamefont {Paul}},\ and\ \bibinfo {author} {\bibfnamefont {A.~D.}\ \bibnamefont {Hillier}},\ }\bibfield  {title} {\enquote {\bibinfo {title} {Using negative muons as a probe for depth profiling silver roman coinage},}\ }\href {https://doi.org/10.3390/heritage2010028} {\bibfield  {journal} {\bibinfo  {journal} {Heritage}\ }\textbf {\bibinfo {volume} {2}},\ \bibinfo {pages} {400--407} (\bibinfo {year} {2019})}\BibitemShut {NoStop}%
\bibitem [{\citenamefont {Ninomiya}\ \emph {et~al.}(2015)\citenamefont {Ninomiya}, \citenamefont {Kubo}, \citenamefont {Nagatomo}, \citenamefont {Higemoto}, \citenamefont {Ito}, \citenamefont {Kawamura}, \citenamefont {Strasser}, \citenamefont {Shimomura}, \citenamefont {Miyake}, \citenamefont {Suzuki}, \citenamefont {Kobayashi}, \citenamefont {Sakamoto}, \citenamefont {Shinohara},\ and\ \citenamefont {Saito}}]{doi:10.1021/acs.analchem.5b01169}%
  \BibitemOpen
  \bibfield  {author} {\bibinfo {author} {\bibfnamefont {K.}~\bibnamefont {Ninomiya}}, \bibinfo {author} {\bibfnamefont {M.~K.}\ \bibnamefont {Kubo}}, \bibinfo {author} {\bibfnamefont {T.}~\bibnamefont {Nagatomo}}, \bibinfo {author} {\bibfnamefont {W.}~\bibnamefont {Higemoto}}, \bibinfo {author} {\bibfnamefont {T.~U.}\ \bibnamefont {Ito}}, \bibinfo {author} {\bibfnamefont {N.}~\bibnamefont {Kawamura}}, \bibinfo {author} {\bibfnamefont {P.}~\bibnamefont {Strasser}}, \bibinfo {author} {\bibfnamefont {K.}~\bibnamefont {Shimomura}}, \bibinfo {author} {\bibfnamefont {Y.}~\bibnamefont {Miyake}}, \bibinfo {author} {\bibfnamefont {T.}~\bibnamefont {Suzuki}}, \bibinfo {author} {\bibfnamefont {Y.}~\bibnamefont {Kobayashi}}, \bibinfo {author} {\bibfnamefont {S.}~\bibnamefont {Sakamoto}}, \bibinfo {author} {\bibfnamefont {A.}~\bibnamefont {Shinohara}},\ and\ \bibinfo {author} {\bibfnamefont {T.}~\bibnamefont {Saito}},\ }\bibfield  {title} {\enquote {\bibinfo {title} {Nondestructive elemental depth-profiling analysis by
  muonic x-ray measurement},}\ }\href {https://doi.org/10.1021/acs.analchem.5b01169} {\bibfield  {journal} {\bibinfo  {journal} {Analytical Chemistry}\ }\textbf {\bibinfo {volume} {87}},\ \bibinfo {pages} {4597--4600} (\bibinfo {year} {2015})},\ \bibinfo {note} {pMID: 25901421},\ \Eprint {https://arxiv.org/abs/https://doi.org/10.1021/acs.analchem.5b01169} {https://doi.org/10.1021/acs.analchem.5b01169} \BibitemShut {NoStop}%
\bibitem [{\citenamefont {Hillier}\ \emph {et~al.}()\citenamefont {Hillier}, \citenamefont {Ishida}, \citenamefont {Seller}, \citenamefont {Veale},\ and\ \citenamefont {Wilson}}]{doi:10.7566/JPSCP.21.011042}%
  \BibitemOpen
  \bibfield  {author} {\bibinfo {author} {\bibfnamefont {A.}~\bibnamefont {Hillier}}, \bibinfo {author} {\bibfnamefont {K.}~\bibnamefont {Ishida}}, \bibinfo {author} {\bibfnamefont {P.}~\bibnamefont {Seller}}, \bibinfo {author} {\bibfnamefont {M.~C.}\ \bibnamefont {Veale}},\ and\ \bibinfo {author} {\bibfnamefont {M.~D.}\ \bibnamefont {Wilson}},\ }\enquote {\bibinfo {title} {Element specific imaging using muonic x-rays},}\ in\ \href {https://doi.org/10.7566/JPSCP.21.011042} {\emph {\bibinfo {booktitle} {Proceedings of the 14th International Conference on Muon Spin Rotation, Relaxation and Resonance (uSR2017)}}},\ \Eprint {https://arxiv.org/abs/https://journals.jps.jp/doi/pdf/10.7566/JPSCP.21.011042} {https://journals.jps.jp/doi/pdf/10.7566/JPSCP.21.011042} \BibitemShut {NoStop}%
\bibitem [{\citenamefont {Katsuragawa}\ \emph {et~al.}(2018)\citenamefont {Katsuragawa}, \citenamefont {Tampo}, \citenamefont {Hamada}, \citenamefont {Harayama}, \citenamefont {Miyake}, \citenamefont {Oshita}, \citenamefont {Sato}, \citenamefont {Takahashi}, \citenamefont {Takeda}, \citenamefont {Watanabe},\ and\ \citenamefont {Yabu}}]{KATSURAGAWA2018140}%
  \BibitemOpen
  \bibfield  {author} {\bibinfo {author} {\bibfnamefont {M.}~\bibnamefont {Katsuragawa}}, \bibinfo {author} {\bibfnamefont {M.}~\bibnamefont {Tampo}}, \bibinfo {author} {\bibfnamefont {K.}~\bibnamefont {Hamada}}, \bibinfo {author} {\bibfnamefont {A.}~\bibnamefont {Harayama}}, \bibinfo {author} {\bibfnamefont {Y.}~\bibnamefont {Miyake}}, \bibinfo {author} {\bibfnamefont {S.}~\bibnamefont {Oshita}}, \bibinfo {author} {\bibfnamefont {G.}~\bibnamefont {Sato}}, \bibinfo {author} {\bibfnamefont {T.}~\bibnamefont {Takahashi}}, \bibinfo {author} {\bibfnamefont {S.}~\bibnamefont {Takeda}}, \bibinfo {author} {\bibfnamefont {S.}~\bibnamefont {Watanabe}},\ and\ \bibinfo {author} {\bibfnamefont {G.}~\bibnamefont {Yabu}},\ }\bibfield  {title} {\enquote {\bibinfo {title} {A compact imaging system with a cdte double-sided strip detector for non-destructive analysis using negative muonic x-rays},}\ }\href {https://doi.org/https://doi.org/10.1016/j.nima.2017.11.004} {\bibfield  {journal} {\bibinfo  {journal} {Nuclear
  Instruments and Methods in Physics Research Section A: Accelerators, Spectrometers, Detectors and Associated Equipment}\ }\textbf {\bibinfo {volume} {912}},\ \bibinfo {pages} {140--143} (\bibinfo {year} {2018})},\ \bibinfo {note} {new Developments In Photodetection 2017}\BibitemShut {NoStop}%
\bibitem [{\citenamefont {Chiu}\ \emph {et~al.}(2022)\citenamefont {Chiu}, \citenamefont {Takeda}, \citenamefont {Kajino}, \citenamefont {Shinohara}, \citenamefont {Katsuragawa}, \citenamefont {Nagasawa}, \citenamefont {Tomaru}, \citenamefont {Yabu}, \citenamefont {Takahashi}, \citenamefont {Watanabe}, \citenamefont {Takeshita}, \citenamefont {Miyake},\ and\ \citenamefont {Ninomiya}}]{Chiu2022}%
  \BibitemOpen
  \bibfield  {author} {\bibinfo {author} {\bibfnamefont {I.-H.}\ \bibnamefont {Chiu}}, \bibinfo {author} {\bibfnamefont {S.}~\bibnamefont {Takeda}}, \bibinfo {author} {\bibfnamefont {M.}~\bibnamefont {Kajino}}, \bibinfo {author} {\bibfnamefont {A.}~\bibnamefont {Shinohara}}, \bibinfo {author} {\bibfnamefont {M.}~\bibnamefont {Katsuragawa}}, \bibinfo {author} {\bibfnamefont {S.}~\bibnamefont {Nagasawa}}, \bibinfo {author} {\bibfnamefont {R.}~\bibnamefont {Tomaru}}, \bibinfo {author} {\bibfnamefont {G.}~\bibnamefont {Yabu}}, \bibinfo {author} {\bibfnamefont {T.}~\bibnamefont {Takahashi}}, \bibinfo {author} {\bibfnamefont {S.}~\bibnamefont {Watanabe}}, \bibinfo {author} {\bibfnamefont {S.}~\bibnamefont {Takeshita}}, \bibinfo {author} {\bibfnamefont {Y.}~\bibnamefont {Miyake}},\ and\ \bibinfo {author} {\bibfnamefont {K.}~\bibnamefont {Ninomiya}},\ }\bibfield  {title} {\enquote {\bibinfo {title} {Non-destructive 3d imaging method using muonic x-rays and a cdte double-sided strip detector},}\ }\href
  {https://doi.org/10.1038/s41598-022-09137-5} {\bibfield  {journal} {\bibinfo  {journal} {Scientific Reports}\ }\textbf {\bibinfo {volume} {12}},\ \bibinfo {pages} {5261} (\bibinfo {year} {2022})}\BibitemShut {NoStop}%
\bibitem [{\citenamefont {Bao}\ \emph {et~al.}(2023)\citenamefont {Bao}, \citenamefont {Chen}, \citenamefont {Chen}, \citenamefont {Cheng}, \citenamefont {Deng}, \citenamefont {Fan}, \citenamefont {Guo}, \citenamefont {He}, \citenamefont {Hu}, \citenamefont {Li}, \citenamefont {Li}, \citenamefont {Liang}, \citenamefont {Liu}, \citenamefont {Lv}, \citenamefont {Pan}, \citenamefont {Tan}, \citenamefont {Vassilopoulos}, \citenamefont {Wu}, \citenamefont {Yang},\ and\ \citenamefont {Zhang}}]{Bao_2023}%
  \BibitemOpen
  \bibfield  {author} {\bibinfo {author} {\bibfnamefont {Y.}~\bibnamefont {Bao}}, \bibinfo {author} {\bibfnamefont {J.}~\bibnamefont {Chen}}, \bibinfo {author} {\bibfnamefont {C.}~\bibnamefont {Chen}}, \bibinfo {author} {\bibfnamefont {H.}~\bibnamefont {Cheng}}, \bibinfo {author} {\bibfnamefont {C.}~\bibnamefont {Deng}}, \bibinfo {author} {\bibfnamefont {R.}~\bibnamefont {Fan}}, \bibinfo {author} {\bibfnamefont {Y.}~\bibnamefont {Guo}}, \bibinfo {author} {\bibfnamefont {N.}~\bibnamefont {He}}, \bibinfo {author} {\bibfnamefont {H.}~\bibnamefont {Hu}}, \bibinfo {author} {\bibfnamefont {Q.}~\bibnamefont {Li}}, \bibinfo {author} {\bibfnamefont {Y.}~\bibnamefont {Li}}, \bibinfo {author} {\bibfnamefont {H.}~\bibnamefont {Liang}}, \bibinfo {author} {\bibfnamefont {L.}~\bibnamefont {Liu}}, \bibinfo {author} {\bibfnamefont {Y.}~\bibnamefont {Lv}}, \bibinfo {author} {\bibfnamefont {Z.}~\bibnamefont {Pan}}, \bibinfo {author} {\bibfnamefont {Z.}~\bibnamefont {Tan}}, \bibinfo {author} {\bibfnamefont {N.}~\bibnamefont
  {Vassilopoulos}}, \bibinfo {author} {\bibfnamefont {Y.}~\bibnamefont {Wu}}, \bibinfo {author} {\bibfnamefont {T.}~\bibnamefont {Yang}},\ and\ \bibinfo {author} {\bibfnamefont {G.}~\bibnamefont {Zhang}},\ }\bibfield  {title} {\enquote {\bibinfo {title} {Progress report on muon source project at csns},}\ }\href {https://doi.org/10.1088/1742-6596/2462/1/012034} {\bibfield  {journal} {\bibinfo  {journal} {Journal of Physics: Conference Series}\ }\textbf {\bibinfo {volume} {2462}},\ \bibinfo {pages} {012034} (\bibinfo {year} {2023})}\BibitemShut {NoStop}%
\bibitem [{\citenamefont {Cai}\ \emph {et~al.}(2024)\citenamefont {Cai}, \citenamefont {He}, \citenamefont {Liu}, \citenamefont {Jia}, \citenamefont {Qin}, \citenamefont {Wang}, \citenamefont {Wang}, \citenamefont {Zhao}, \citenamefont {Pu}, \citenamefont {Niu}, \citenamefont {Chen}, \citenamefont {Sun}, \citenamefont {Zhao},\ and\ \citenamefont {Zhan}}]{PhysRevAccelBeams.27.023403}%
  \BibitemOpen
  \bibfield  {author} {\bibinfo {author} {\bibfnamefont {H.-J.}\ \bibnamefont {Cai}}, \bibinfo {author} {\bibfnamefont {Y.}~\bibnamefont {He}}, \bibinfo {author} {\bibfnamefont {S.}~\bibnamefont {Liu}}, \bibinfo {author} {\bibfnamefont {H.}~\bibnamefont {Jia}}, \bibinfo {author} {\bibfnamefont {Y.}~\bibnamefont {Qin}}, \bibinfo {author} {\bibfnamefont {Z.}~\bibnamefont {Wang}}, \bibinfo {author} {\bibfnamefont {F.}~\bibnamefont {Wang}}, \bibinfo {author} {\bibfnamefont {L.}~\bibnamefont {Zhao}}, \bibinfo {author} {\bibfnamefont {N.}~\bibnamefont {Pu}}, \bibinfo {author} {\bibfnamefont {J.}~\bibnamefont {Niu}}, \bibinfo {author} {\bibfnamefont {L.}~\bibnamefont {Chen}}, \bibinfo {author} {\bibfnamefont {Z.}~\bibnamefont {Sun}}, \bibinfo {author} {\bibfnamefont {H.}~\bibnamefont {Zhao}},\ and\ \bibinfo {author} {\bibfnamefont {W.}~\bibnamefont {Zhan}},\ }\bibfield  {title} {\enquote {\bibinfo {title} {Towards a high-intensity muon source},}\ }\href {https://doi.org/10.1103/PhysRevAccelBeams.27.023403}
  {\bibfield  {journal} {\bibinfo  {journal} {Phys. Rev. Accel. Beams}\ }\textbf {\bibinfo {volume} {27}},\ \bibinfo {pages} {023403} (\bibinfo {year} {2024})}\BibitemShut {NoStop}%
\bibitem [{\citenamefont {Eaton}(1992)}]{Eaton1992}%
  \BibitemOpen
  \bibfield  {author} {\bibinfo {author} {\bibfnamefont {G.~H.}\ \bibnamefont {Eaton}},\ }\bibfield  {title} {\enquote {\bibinfo {title} {The isis pulsed muon facility},}\ }\href {https://doi.org/10.1007/BF02426802} {\bibfield  {journal} {\bibinfo  {journal} {Zeitschrift f{\"u}r Physik C Particles and Fields}\ }\textbf {\bibinfo {volume} {56}},\ \bibinfo {pages} {S232--S239} (\bibinfo {year} {1992})}\BibitemShut {NoStop}%
\bibitem [{\citenamefont {Miyake}\ \emph {et~al.}(2009)\citenamefont {Miyake}, \citenamefont {Nishiyama}, \citenamefont {Kawamura}, \citenamefont {Strasser}, \citenamefont {Makimura}, \citenamefont {Koda}, \citenamefont {Shimomura}, \citenamefont {Fujimori}, \citenamefont {Nakahara}, \citenamefont {Kadono}, \citenamefont {Kato}, \citenamefont {Takeshita}, \citenamefont {Higemoto}, \citenamefont {Ishida}, \citenamefont {Matsuzaki}, \citenamefont {Matsuda},\ and\ \citenamefont {Nagamine}}]{MIYAKE200922}%
  \BibitemOpen
  \bibfield  {author} {\bibinfo {author} {\bibfnamefont {Y.}~\bibnamefont {Miyake}}, \bibinfo {author} {\bibfnamefont {K.}~\bibnamefont {Nishiyama}}, \bibinfo {author} {\bibfnamefont {N.}~\bibnamefont {Kawamura}}, \bibinfo {author} {\bibfnamefont {P.}~\bibnamefont {Strasser}}, \bibinfo {author} {\bibfnamefont {S.}~\bibnamefont {Makimura}}, \bibinfo {author} {\bibfnamefont {A.}~\bibnamefont {Koda}}, \bibinfo {author} {\bibfnamefont {K.}~\bibnamefont {Shimomura}}, \bibinfo {author} {\bibfnamefont {H.}~\bibnamefont {Fujimori}}, \bibinfo {author} {\bibfnamefont {K.}~\bibnamefont {Nakahara}}, \bibinfo {author} {\bibfnamefont {R.}~\bibnamefont {Kadono}}, \bibinfo {author} {\bibfnamefont {M.}~\bibnamefont {Kato}}, \bibinfo {author} {\bibfnamefont {S.}~\bibnamefont {Takeshita}}, \bibinfo {author} {\bibfnamefont {W.}~\bibnamefont {Higemoto}}, \bibinfo {author} {\bibfnamefont {K.}~\bibnamefont {Ishida}}, \bibinfo {author} {\bibfnamefont {T.}~\bibnamefont {Matsuzaki}}, \bibinfo {author} {\bibfnamefont {Y.}~\bibnamefont
  {Matsuda}},\ and\ \bibinfo {author} {\bibfnamefont {K.}~\bibnamefont {Nagamine}},\ }\bibfield  {title} {\enquote {\bibinfo {title} {J-parc muon source, muse},}\ }\href {https://doi.org/https://doi.org/10.1016/j.nima.2008.11.016} {\bibfield  {journal} {\bibinfo  {journal} {Nuclear Instruments and Methods in Physics Research Section A: Accelerators, Spectrometers, Detectors and Associated Equipment}\ }\textbf {\bibinfo {volume} {600}},\ \bibinfo {pages} {22--24} (\bibinfo {year} {2009})}\BibitemShut {NoStop}%
\bibitem [{\citenamefont {Gottesman}\ and\ \citenamefont {Fenimore}(1989)}]{Gottesman:89}%
  \BibitemOpen
  \bibfield  {author} {\bibinfo {author} {\bibfnamefont {S.~R.}\ \bibnamefont {Gottesman}}\ and\ \bibinfo {author} {\bibfnamefont {E.~E.}\ \bibnamefont {Fenimore}},\ }\bibfield  {title} {\enquote {\bibinfo {title} {New family of binary arrays for coded aperture imaging},}\ }\href {https://doi.org/10.1364/AO.28.004344} {\bibfield  {journal} {\bibinfo  {journal} {Appl. Opt.}\ }\textbf {\bibinfo {volume} {28}},\ \bibinfo {pages} {4344--4352} (\bibinfo {year} {1989})}\BibitemShut {NoStop}%
\bibitem [{\citenamefont {Lin}\ \emph {et~al.}(2022)\citenamefont {Lin}, \citenamefont {Pan}, \citenamefont {Wang}, \citenamefont {He}, \citenamefont {Dong}, \citenamefont {Liu}, \citenamefont {Zhang},\ and\ \citenamefont {Ye}}]{LIN2022166783}%
  \BibitemOpen
  \bibfield  {author} {\bibinfo {author} {\bibfnamefont {Z.}~\bibnamefont {Lin}}, \bibinfo {author} {\bibfnamefont {Z.}~\bibnamefont {Pan}}, \bibinfo {author} {\bibfnamefont {Z.}~\bibnamefont {Wang}}, \bibinfo {author} {\bibfnamefont {Z.}~\bibnamefont {He}}, \bibinfo {author} {\bibfnamefont {J.}~\bibnamefont {Dong}}, \bibinfo {author} {\bibfnamefont {J.}~\bibnamefont {Liu}}, \bibinfo {author} {\bibfnamefont {H.}~\bibnamefont {Zhang}},\ and\ \bibinfo {author} {\bibfnamefont {B.}~\bibnamefont {Ye}},\ }\bibfield  {title} {\enquote {\bibinfo {title} {A study of the feasibility of coded aperture imaging technique for elemental analysis by muonic x-ray measurements based on geant4 simulations},}\ }\href {https://doi.org/https://doi.org/10.1016/j.nima.2022.166783} {\bibfield  {journal} {\bibinfo  {journal} {Nuclear Instruments and Methods in Physics Research Section A: Accelerators, Spectrometers, Detectors and Associated Equipment}\ }\textbf {\bibinfo {volume} {1034}},\ \bibinfo {pages} {166783} (\bibinfo {year}
  {2022})}\BibitemShut {NoStop}%
\bibitem [{\citenamefont {Yu}\ \emph {et~al.}(2024)\citenamefont {Yu}, \citenamefont {Li}, \citenamefont {Dai}, \citenamefont {Zhang}, \citenamefont {Chen}, \citenamefont {Zhong}, \citenamefont {Wang}, \citenamefont {Xia}, \citenamefont {Cao}, \citenamefont {Zhou},\ and\ \citenamefont {Ruan}}]{10.1063/5.0180056}%
  \BibitemOpen
  \bibfield  {author} {\bibinfo {author} {\bibfnamefont {J.}~\bibnamefont {Yu}}, \bibinfo {author} {\bibfnamefont {D.}~\bibnamefont {Li}}, \bibinfo {author} {\bibfnamefont {Y.}~\bibnamefont {Dai}}, \bibinfo {author} {\bibfnamefont {C.}~\bibnamefont {Zhang}}, \bibinfo {author} {\bibfnamefont {W.}~\bibnamefont {Chen}}, \bibinfo {author} {\bibfnamefont {J.}~\bibnamefont {Zhong}}, \bibinfo {author} {\bibfnamefont {X.}~\bibnamefont {Wang}}, \bibinfo {author} {\bibfnamefont {R.}~\bibnamefont {Xia}}, \bibinfo {author} {\bibfnamefont {L.}~\bibnamefont {Cao}}, \bibinfo {author} {\bibfnamefont {C.}~\bibnamefont {Zhou}},\ and\ \bibinfo {author} {\bibfnamefont {S.}~\bibnamefont {Ruan}},\ }\bibfield  {title} {\enquote {\bibinfo {title} {{Size characterization of x-ray tube source with sphere encoded imaging method}},}\ }\href {https://doi.org/10.1063/5.0180056} {\bibfield  {journal} {\bibinfo  {journal} {Review of Scientific Instruments}\ }\textbf {\bibinfo {volume} {95}},\ \bibinfo {pages} {013102} (\bibinfo {year}
  {2024})},\ \Eprint {https://arxiv.org/abs/https://pubs.aip.org/aip/rsi/article-pdf/doi/10.1063/5.0180056/18701441/013102\_1\_5.0180056.pdf} {https://pubs.aip.org/aip/rsi/article-pdf/doi/10.1063/5.0180056/18701441/013102\_1\_5.0180056.pdf} \BibitemShut {NoStop}%
\bibitem [{\citenamefont {Kozioziemski}\ \emph {et~al.}(2023)\citenamefont {Kozioziemski}, \citenamefont {Bachmann}, \citenamefont {Do},\ and\ \citenamefont {Tommasini}}]{10.1063/5.0130689}%
  \BibitemOpen
  \bibfield  {author} {\bibinfo {author} {\bibfnamefont {B.}~\bibnamefont {Kozioziemski}}, \bibinfo {author} {\bibfnamefont {B.}~\bibnamefont {Bachmann}}, \bibinfo {author} {\bibfnamefont {A.}~\bibnamefont {Do}},\ and\ \bibinfo {author} {\bibfnamefont {R.}~\bibnamefont {Tommasini}},\ }\bibfield  {title} {\enquote {\bibinfo {title} {{X-ray imaging methods for high-energy density physics applications}},}\ }\href {https://doi.org/10.1063/5.0130689} {\bibfield  {journal} {\bibinfo  {journal} {Review of Scientific Instruments}\ }\textbf {\bibinfo {volume} {94}},\ \bibinfo {pages} {041102} (\bibinfo {year} {2023})},\ \Eprint {https://arxiv.org/abs/https://pubs.aip.org/aip/rsi/article-pdf/doi/10.1063/5.0130689/17106000/041102\_1\_5.0130689.pdf} {https://pubs.aip.org/aip/rsi/article-pdf/doi/10.1063/5.0130689/17106000/041102\_1\_5.0130689.pdf} \BibitemShut {NoStop}%
\bibitem [{\citenamefont {Bachmann}\ \emph {et~al.}(2016)\citenamefont {Bachmann}, \citenamefont {Hilsabeck}, \citenamefont {Field}, \citenamefont {Masters}, \citenamefont {Reed}, \citenamefont {Pardini}, \citenamefont {Rygg}, \citenamefont {Alexander}, \citenamefont {Benedetti}, \citenamefont {Döppner}, \citenamefont {Forsman}, \citenamefont {Izumi}, \citenamefont {LePape}, \citenamefont {Ma}, \citenamefont {MacPhee}, \citenamefont {Nagel}, \citenamefont {Patel}, \citenamefont {Spears},\ and\ \citenamefont {Landen}}]{10.1063/1.4959161}%
  \BibitemOpen
  \bibfield  {author} {\bibinfo {author} {\bibfnamefont {B.}~\bibnamefont {Bachmann}}, \bibinfo {author} {\bibfnamefont {T.}~\bibnamefont {Hilsabeck}}, \bibinfo {author} {\bibfnamefont {J.}~\bibnamefont {Field}}, \bibinfo {author} {\bibfnamefont {N.}~\bibnamefont {Masters}}, \bibinfo {author} {\bibfnamefont {C.}~\bibnamefont {Reed}}, \bibinfo {author} {\bibfnamefont {T.}~\bibnamefont {Pardini}}, \bibinfo {author} {\bibfnamefont {J.~R.}\ \bibnamefont {Rygg}}, \bibinfo {author} {\bibfnamefont {N.}~\bibnamefont {Alexander}}, \bibinfo {author} {\bibfnamefont {L.~R.}\ \bibnamefont {Benedetti}}, \bibinfo {author} {\bibfnamefont {T.}~\bibnamefont {Döppner}}, \bibinfo {author} {\bibfnamefont {A.}~\bibnamefont {Forsman}}, \bibinfo {author} {\bibfnamefont {N.}~\bibnamefont {Izumi}}, \bibinfo {author} {\bibfnamefont {S.}~\bibnamefont {LePape}}, \bibinfo {author} {\bibfnamefont {T.}~\bibnamefont {Ma}}, \bibinfo {author} {\bibfnamefont {A.~G.}\ \bibnamefont {MacPhee}}, \bibinfo {author} {\bibfnamefont {S.}~\bibnamefont
  {Nagel}}, \bibinfo {author} {\bibfnamefont {P.}~\bibnamefont {Patel}}, \bibinfo {author} {\bibfnamefont {B.}~\bibnamefont {Spears}},\ and\ \bibinfo {author} {\bibfnamefont {O.~L.}\ \bibnamefont {Landen}},\ }\bibfield  {title} {\enquote {\bibinfo {title} {{Resolving hot spot microstructure using x-ray penumbral imaging (invited)}},}\ }\href {https://doi.org/10.1063/1.4959161} {\bibfield  {journal} {\bibinfo  {journal} {Review of Scientific Instruments}\ }\textbf {\bibinfo {volume} {87}},\ \bibinfo {pages} {11E201} (\bibinfo {year} {2016})},\ \Eprint {https://arxiv.org/abs/https://pubs.aip.org/aip/rsi/article-pdf/doi/10.1063/1.4959161/16053294/11e201\_1\_online.pdf} {https://pubs.aip.org/aip/rsi/article-pdf/doi/10.1063/1.4959161/16053294/11e201\_1\_online.pdf} \BibitemShut {NoStop}%
\end{thebibliography}%

\end{document}